\documentclass{JHEP3}

\title{On deformed gauge theories and their string/M-theory duals}

\author{Emiliano Imeroni\\
Physique Th\'eorique et Math\'ematique, Universit\'e Libre de Bruxelles\\
\& International Solvay Institutes, CP 231, 1050 Bruxelles, Belgium\\
E-mail: \email{eimeroni@ulb.ac.be}}

\abstract{We present general formulae for the TsT transformation (T-duality, shift, T-duality) of type II string backgrounds and open string boundary conditions. The TsT transformation provides a systematic procedure to find string theory duals of gauge theories with deformed products of fields in the lagrangian, and the duals can be analyzed by using transformed D-brane probes. As examples illustrating some features of the deformed theories, we consider the known backgrounds dual to non-commutative, dipole and $\beta$-deformed \Ne{4} Super Yang-Mills as well as new backgrounds dual to deformations of the recently proposed \Ne{6} Chern-Simons-matter theory living on multiple M2-branes on an orbifold.}


\preprint{}

\usepackage{amsmath}
\usepackage{graphicx}

\DeclareMathOperator{\tr}{Tr} % Trace
\newcommand{\Ne}[1]{\ensuremath{\mathcal{N}={#1}}} % Supersymmetry: N = ?
\newcommand{\ls}{\ensuremath{{l_s}}} % String length
\newcommand{\lp}{\ensuremath{{l_p}}} % Planck length
\newcommand{\gs}{\ensuremath{{g_s}}} % String coupling
\newcommand{\ads}{{\ensuremath{AdS_5\times S^5}}} % AdS_5 x S^5
\newcommand{\sk}{{\ensuremath{S^7 / \mathbb{Z}_k}}} % S^7 / Z_k
\newcommand{\adsk}{{\ensuremath{AdS_4\times \sk}}} % AdS_4 x S^7 / Z_k
\newcommand{\cp}{{\ensuremath{\mathbb{CP}^3}}} % CP^3
\newcommand{\hd}[1]{\ensuremath{\phantom{ }^{\star_{#1}}}} % Hodge dual in d dimensions
\newcommand{\tst}[3]{\ensuremath{( #1 , #2 )^{\text{TsT}}_{ #3 }}}

\newcommand{\cZ}[2]{{\ensuremath{\mathbb{C}^{#1}/\mathbb{Z}_{#2}}}} % C^? / Z_?
\newcommand{\tvarphi}{{\tilde{\varphi}}}

\newcommand{\hgamma}{{\hat{\gamma}}}

\newcommand{\M}{\mathcal{M}}

\begin{document}

\section{Introduction}

Deformed field theories, that arise from a new definition of the product of fields in the lagrangian, constitute an interesting generalization of the gauge/gravity correspondence~\cite{Maldacena:1998re,Gubser:1998bc,Witten:1998qj}. This is also due to the fact that, on the string theory side, there is a systematic procedure called the ``TsT transformation'' (T-duality, shift, T-duality)~\cite{Lunin:2005jy} that can be applied in order to derive the dual supergravity solutions as well as analyze D-brane probes embedded in the corresponding backgrounds.

These deformed gauge theories are obtained from ordinary theories by replacing the ordinary point-wise product of two fields in the lagrangian by the  ``star'' product:
\begin{equation}\label{star}
	fg \to f \star g = e^{i \pi \gamma (p^f_1 p^g_2 - p^f_2 p^g_1)} f g\,,
\end{equation}
where $p_i$ are appropriately chosen charges of the fields $f$ and $g$ and $\gamma$ is a real parameter. For instance, if we take these charges to be momenta along two compactified space-time directions, the product~\eqref{star} will reduce to the usual Moyal product and the gauge theory after the deformation will be defined on a non-commutative two-torus~\cite{Connes:1987ue,Connes:1997cr,Seiberg:1999vs}. Different choices for the charges $p_i$ can lead to different, and often less ``exotic'', deformations.

In fact, deformations of the type~\eqref{star} can be seen in a unified framework as emerging from deformations of ordinary Yang-Mills theory by higher-dimensional (but not necessarily irrelevant) gauge-invariant operators. It is this perspective that allowed Lunin and Maldacena to find the string dual of the deformation~\cite{Lunin:2005jy}. Assuming that the gravity dual of the original, undeformed, gauge theory has a two-torus isometry, the gravity description of the deformation just consists in the following $SL(2,\mathbb{R})$ redefinition of the complex structure of the two-torus:
\begin{equation}\label{SL2R}
	\tau \to \tau_{\gamma} = \frac{\tau}{ 1 + \gamma \tau }\,.
\end{equation}
This transformation can be seen as a solution-generating technique to obtain the gravity duals of the deformed gauge theories~\eqref{star}. In particular, on a supergravity solution of a type II theory, it reduces to a simple ``TsT'' transformation. If we parameterize the two torus by $(\varphi^1,\varphi^2)$, the transformation consists of a T-duality along $\varphi^1$, followed by a shift $\varphi^2 \to \varphi^2 + \gamma \varphi^1$ in the T-dual background, and finally by another T-duality  along $\varphi^1$.

Gauge theories with deformed products of fields and their string theory duals, obtained with or without the knowledge of the transformation~\eqref{SL2R}, have been thoroughly studied in the literature. The purpose of this paper is to present general formulae for the TsT transformation of any type II background, including the transformation of the string world-sheet coordinate fields (the study of which was initiated in~\cite{Frolov:2005dj}) and of the corresponding open string boundary conditions. In particular, we will give a general perspective on how D-brane probes behave under the transformation, which we think is particularly relevant as D-brane probes often provide an invaluable ``bridge'' between the gauge and string sides of the correspondence.

A quick reference for the reader interested in our general results is the following: the formulae for the TsT transformation of a type II background are given in equations~\eqref{NSNSTsT}, \eqref{M} and \eqref{RRTsT} of section~\ref{s:closed}, while formulae for the TsT of world-sheet fields are given in equation~\eqref{TsTwv} of section~\ref{s:open}.

Our results can be immediately applied to a plethora of situations, in particular in contexts where the TsT transformation has been shown to be useful. The most well-known example is probably the one of exactly marginal deformations of (super)conformal gauge theories, such as the deformations of \Ne{4} Super Yang-Mills theory considered in~\cite{Leigh:1995ep}, whose gravity dual (for the so-called $\beta$-deformation) was derived in~\cite{Lunin:2005jy}. Another example, as we have said in the beginning, is provided by gravity duals of non-commutative gauge theories, such as the ones in~\cite{Hashimoto:1999ut,Maldacena:1999mh}.

Another often studied case is the one of ``dipole'' theories~\cite{Bergman:2000cw,Dasgupta:2000ry,Bergman:2001rw,Alishahiha:2003ru}, which are non-local theories living on an ordinary commutative space. Besides being interesting in themselves, dipole deformations have been shown to be useful also in the context of ordinary confining \Ne{1} gauge theories realized on D-branes wrapped on supersymmetric cycles of Calabi-Yau manifolds (such as in~\cite{Maldacena:2000yy}), where the deformation helps with disentangling gauge theory effect from spurious effects due to the Kaluza-Klein modes on the cycle~\cite{Gursoy:2005cn}. The general TsT formulae we present here can also for instance be used~\cite{Imeroni:2008??} in the case of solutions of the combined type II supergravity and D-brane world-volume actions, such as the solutions dual to \Ne{1} SQCD with a large number of fundamental flavors~\cite{Casero:2006pt,Casero:2007jj}.

Dipole-type TsT deformations have also recently been applied in the context of non-relativistic AdS/CFT~\cite{Herzog:2008wg,Maldacena:2008wh,Adams:2008wt}, where the transformation involves a light-cone direction of the gravity solution.

Besides presenting general formulae for TsT transformations, we have chosen a couple of examples that we hope will shed light on some of the relevant features of the deformed theories and their gravity duals. Both the examples we consider start with superconformal gauge theories, of which we consider non-commutative, dipole and $\beta$-deformations. In the first example, we start from \Ne{4} Super Yang-Mills and its well-known \ads{} gravity dual in order to derive, in a unified perspective, known type IIB supergravity backgrounds dual to its non-commutative, dipole and (non)supersymmetric $\beta$-deformations. In particular, in the case of the $\beta$-deformation we will make use of transformed D-brane probes to see the effect of the deformation on the vacua of the theory.

By studying the second example we will instead derive new type IIA and eleven-dimensional supergravity solutions,  which are dual to deformations of the recently proposed \Ne{6} superconformal Chern-Simons-matter theory in three-dimensions that has been conjectured to live on $N$ M-theory M2-branes at a $\cZ{4}{k}$ orbifold singularity~\cite{Aharony:2008ug}. Once again, in the case of the $\beta$-deformation we will show some interesting results arising from the study of D-branes in the transformed background.

This paper is organized as follows. We start in section~\ref{s:gauge} by reviewing the introduction of the deformed ``star'' product in gauge theory lagrangians and its consequences on the string theory side of the duality. In sections~\ref{s:closed} and~\ref{s:open} we present general formulae for the TsT transformation of type II closed string backgrounds, world-sheet coordinate fields and the corresponding open string boundary conditions, then we proceed in section~\ref{s:branes} by studying how D-branes behave along the transformation of a very simple background. The two final sections are devoted to concrete examples illustrating some features of the transformed theories: in section~\ref{s:SYM}, we consider deformations of \Ne{4} Super Yang-Mills theory in four dimensions and their string duals, while in section~\ref{s:ABJM} we study deformations of the \Ne{6} ABJM theory. We outline our conventions and collect a few useful formulae in the appendix.

\section{Gauge lagrangians with deformed products and their gravity duals}\label{s:gauge}

We want to consider gauge theories with deformed products of fields in the lagrangian, because they will lead us to interesting generalizations of the gauge/string theory correspondence. Given two fields $f$ and $g$, we replace the ordinary point-wise local product $fg$ by a generic ``star'' product, which is defined as:
\begin{equation}\label{star2}
	 f \star g = e^{i \pi \gamma (p^f_1 p^g_2 - p^f_2 p^g_1)} f g\,,
\end{equation}
where the $p_i$ are appropriate conserved charges of the fields $f$ and $g$ (the expression above should be regarded as schematic) and $\gamma$ is an arbitrary real parameter. What are the properties of the deformed theory? Let us consider in more detail three very different cases.

\begin{enumerate}
\item In the first case, we take the $p_i$ charges to be the conserved momenta along two space-time directions, that we think of as being compactified on circles of radius $2 \pi$. In this case, the deformed product~\eqref{star2} becomes:
\begin{equation}\label{Moyal}
\begin{split}
	( f \star g ) (x) &= e^{- i \pi \gamma \left(\frac{\partial}{\partial x^1} \frac{\partial}{\partial y^2}
		- \frac{\partial}{\partial x^2} \frac{\partial}{\partial y^1} \right)} f (x) g (y) \big\rvert_{x=y}\\
		&= f(x) g(x)
			- i \pi \gamma \left( \partial_1 f(x) \partial_2 g(x) - \partial_2 f(x) \partial_1 g(x) \right)
			+ \ldots
\end{split}
\end{equation}
This can be recognized as the appropriate Moyal product for a non-commutative two-torus whose coordinates satisfy:
\begin{equation}
	[ x^1 , x^2 ] = i \theta^{12}\,\qquad
	\theta^{12} = - 2 \pi \gamma\,.
\end{equation}
This deformation then yields a non-commutative theory, which is non-local and breaks Lorentz invariance and causality. Of course we can easily generalize this deformation to obtain a theory living on a larger non-commutative space whose coordinates obey the relations $[ x^i , x^j ] = i \theta^{ij}$.
\item In the second case, let us suppose that there is a global $U(1)$ conserved charge under which the fields have charge $p^M_2 = Q^M$ ($M$ is an index distinguishing the various fields of the theory). If we moreover take the first charge $p_1$ to be the momentum, the star product becomes:
\begin{equation}\label{stardip}
	 ( f \star g ) (x) = e^{\pi \gamma \left(  Q^g \frac{\partial}{\partial x}
	 	- Q^f \frac{\partial}{\partial y} \right)} f (x) g (y) \big\rvert_{x=y}
	 	= f ( x + \pi \gamma Q^g ) g ( x - \pi \gamma Q^f )\,.
\end{equation}
This is called dipole deformation~\cite{Bergman:2000cw,Dasgupta:2000ry,Bergman:2001rw,Alishahiha:2003ru}, and is clearly non-local despite living on a commutative space-time. Here we have shifted a single direction $x$, but we can obtain more general deformations by introducing ``dipole vectors'' $L^{M\mu} = - 2 \pi \gamma Q^M L^{\mu}$ for the various fields spanned by the index $M$, where $L^\mu$ is a constant vector. In this case we can rewrite the product as:
\begin{equation}
	 ( f \star g ) (x) = f \big( x - \tfrac{1}{2} L^g \big) g \big( x+ \tfrac{1}{2} L^f \big)\,.
\end{equation}
\item In the last case, let us suppose there are two global $U(1)$ conserved charges,  $p^M_i = Q^M_i$, so that the star product~\eqref{star2} reduces to:
\begin{equation}\label{starbeta}
	 f \star g = e^{i \pi \gamma (Q^f_1 Q^g_2 - Q^f_2 Q^g_1)} f g\,.
\end{equation}
In this case, we see that the deformation yields an ordinary theory, commutative and local, since the only effect of the deformed product~\eqref{starbeta} is to introduce some phases in the interactions. The theory arising from the product~\eqref{starbeta} is called the $\beta$-deformation~\cite{Leigh:1995ep} and, in the case of superconformal field theories, can be shown to be an exactly marginal deformation.%
\footnote{The general $\beta = \gamma + i \sigma$ parameter of the $\beta$-deformation is complex. In this paper, we will limit ourselves to the case where $\beta$ is a real number, $\beta = \gamma$.}
In the case of supersymmetric theories, one can see the deformation~\eqref{starbeta} as acting directly on the superpotential, as we will show later in examples.
\end{enumerate}

%: Figure: TsT transformation of gauge theories
\FIGURE{\label{f:TsT}
\includegraphics[scale=1]{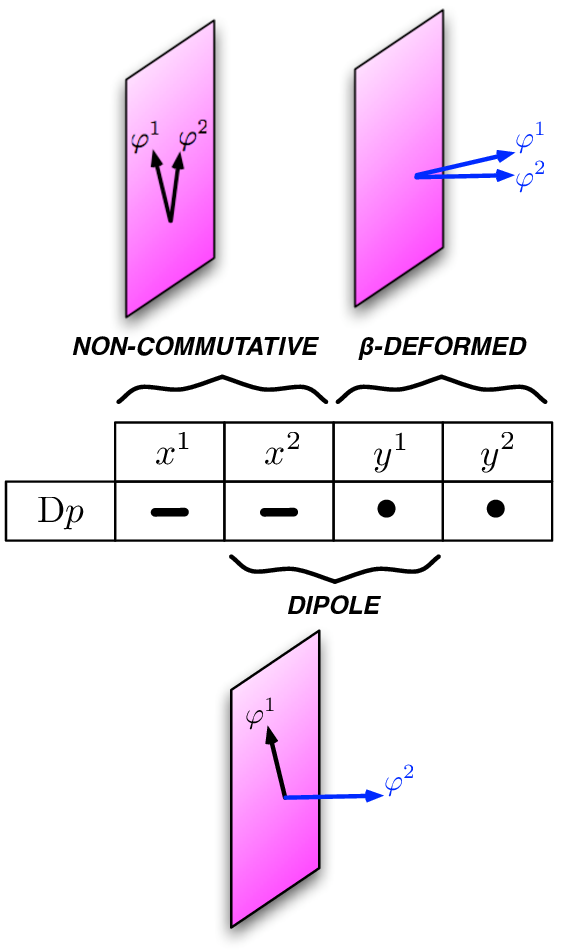}
\caption{The TsT transformation yields different theories when applied to different directions. In the figure, $\,\mathbf{-}\,$ and $\,\bullet\,$ indicate respectively longitudinal and transverse directions to the world-volume of the D$p$-brane that hosts the gauge theory to be deformed.}}

As we mentioned in the introduction, from the point of view of our analysis one of the nicest things about deformations of gauge theories of the general form \eqref{star2} is that they can be seen as arising from ordinary Yang-Mills theory via deformations induced by higher-dimensional (but not necessarily irrelevant, as the example of the exactly marginal $\beta$-deformation shows) gauge invariant operators, so that we can treat them in a unified way. Let us then try to see what these deformations mean from the string theory point of view, when we think of the gauge theories as being the world-volume theories living on D-branes of the type II string. The answer was found by Lunin and Maldacena~\cite{Lunin:2005jy}, who have shown that the deformation amounts to the $SL(2,\mathbb{R})$ transformation
\begin{equation}\label{SL2R2}
	\tau \to \tau_{\gamma} = \frac{\tau}{ 1 + \gamma \tau }
\end{equation}
on the modulus of a two-torus present in the geometry.

The transformation~\eqref{SL2R2} of the $(\varphi^1,\varphi^2)$ two-torus corresponds to a ``TsT'' transformation of the ten-dimensional type II background generated by the D-brane where the gauge theory under consideration lives (or the corresponding near-horizon limit, if we are in the context of the gauge/string duality). The transformation amounts to a T-duality along the isometry direction $\varphi^1$, then a shift $\varphi^2 \to \varphi^2 + \gamma \varphi^1$, and finally to another T-duality along $\varphi^1$. The resulting deformed gauge theory can then have different properties, depending on how the coordinates $\varphi^i$ are related to the world-volume directions of the D-brane. It is clear that there will be three possibilities, as summarized in figure~\ref{f:TsT}:
\begin{enumerate}
\item The torus is entirely part of the world-volume of the D-brane. In this case, the charges that correspond to the torus directions will translate into local momenta of the gauge theory. Applying~\eqref{SL2R2} to the geometry generated by the D-brane will then yield the non-commutative deformation~\eqref{Moyal}.
\item The torus has one direction along the brane and the other transverse to the brane. In this case the transverse direction will correspond to the $U(1)$ global charge and we land on a dipole deformation~\eqref{stardip}.
\item The torus is transverse to the brane. In this case there are $U(1) \times U(1)$ global charges for the fields of the world-volume theory, and the transformation will result in a $\beta$-deformed theory~\eqref{starbeta}.
\end{enumerate}

Finding general formulae for the TsT transformation provides a useful tool to study all of these deformations from the point of view of the gauge/string duality, and this is where our attention turns to in the following section.

\section{TsT transformation: space-time perspective}\label{s:closed}

In this section we present explicit general formulae for the ``TsT transformation'' of a type II closed string background.

We start with type IIA(B) supergravity solution with metric $g_{\mu\nu}$, dilaton $\phi$, antisymmetric tensor field $b_{\mu\nu}$ and modified Ramond-Ramond field strengths $f_p = d c_{p-1} + d b \wedge c_{p-3}$, where $p$ is even in type IIA and odd in type IIB. We define $e_{\mu\nu} = g_{\mu\nu} + b_{\mu\nu}$.

Assume that the coordinates $\varphi^\alpha$ for $\alpha=1,2$ are two commuting $U(1)$ isometries of the solution. Then the TsT transformation along $\varphi^\alpha$ with parameter $\gamma$, that we denote as \tst{\varphi^1}{\varphi^2}{\gamma}, consists of three steps:
\begin{enumerate}
\item Perform a T-duality along $\varphi^1$. The type IIA(B) solution becomes a type IIB(A) solution with T-dual coordinate $\tvarphi^1$.
\item Shift the coordinates in the new solution as follows:
\begin{equation}
	\tvarphi^2 \to \tvarphi^2 + \gamma \tvarphi^1\,,
\end{equation}
where $\gamma$ is an arbitrary real parameter.
\item Perform another T-duality along $\tvarphi^1$, going back to the type IIA(B) theory. We call the final T-dual coordinate $\varphi^1$ again, hoping this does not cause any confusion.
\end{enumerate}  

As discussed in~\cite{Lunin:2005jy}, TsT is a solution-generating technique. The shift may change the periodicities of the angles $\varphi^\alpha$, but the transformation is guaranteed to give a new solution of the supergravity equations of motion. Under certain conditions, the transformation will also not generate any new singularities.

Let us then derive the TsT formulae (general expressions have also been written elsewhere in different forms, see for instance~\cite{CatalOzer:2005mr}). By using the standard T-duality rules, summarized in appendix~\ref{s:conventions}, we can show that the NS-NS fields $E_{\mu\nu} = G_{\mu\nu} + B_{\mu\nu}$ and $\Phi$ of the TsT-transformed solution can be obtained from $e_{\mu\nu}$ and $\phi$ of the undeformed solution as follows:
\begin{equation}\label{NSNSTsT}
\begin{split}
	E_{\mu\nu} &= \M \left\{ e_{\mu\nu}
		- \gamma \left[ \det \begin{pmatrix}
			e_{12} & e_{1 \nu}\\
			e_{\mu 2} & e_{\mu \nu}
			\end{pmatrix}
			- \det \begin{pmatrix}
			e_{21} & e_{2 \nu}\\
			e_{\mu 1} & e_{\mu \nu}
			\end{pmatrix} \right]
		+ \gamma^2 \det \begin{pmatrix}
			e_{11} & e_{12} & e_{1\nu}\\
			e_{21} & e_{22} & e_{2\nu}\\
			e_{\mu 1} & e_{\mu 2} & e_{\mu\nu}
			\end{pmatrix} \right\} \,,\\
	e^{2 \Phi} &= \M \ e^{2 \phi}\,,
\end{split}
\end{equation}
where we have defined the quantity:
\begin{equation}\label{M}
	\M = \left\{ 1 - \gamma \left(e_{12} - e_{21} \right) + \gamma^2 \det \begin{pmatrix}
			e_{11} & e_{12}\\
			e_{21} & e_{22}
			\end{pmatrix} \right\}^{-1}\,.
\end{equation}

Repeating the process in the R-R sector, we find that the easiest way to express the new R-R modified field strengths $\mathcal{F}_p = F_p + H \wedge C_{p-3}$ is by means of the general formula:
\begin{equation}\label{RRTsT}
	\sum_q \mathcal{F}_q \wedge e^{B}
		= \sum_q f_q \wedge e^{b}
		+ \gamma \left[ \sum_q f_q \wedge e^{b} \right]_{[\varphi^1][\varphi^2]}\,,
\end{equation}
where $q$ is even (odd) in type IIA(B), and the anticommuting interior product operation $\cdot_{[y]}$ acts on a $p$-form and gives a $(p-1)$-form whose components are given by~\eqref{omegay}:
\begin{equation}
	(\omega_{p [y]})_{\alpha_1 \dotsm \alpha_{p-1}} = (\omega_p)_{\alpha_1 \dotsm \alpha_{p-1} y}\,.
\end{equation}
Formula~\eqref{RRTsT} must of course be understood as a symbolic expression that is valid degree by degree of the differential forms. More explicitly, for the field strengths appearing in the action of type IIB we have:
\begin{equation}
\begin{split}
	F_1 &= f_1 + \gamma \left[ f_3 + f_1 \wedge b \right]_{[\varphi^1][\varphi^2]}\,,\\
	\mathcal{F}_3 + F_1 \wedge B 
		&= f_3 + f_1 \wedge b + \gamma \left[ f_5 + f_3 \wedge b
			+ \frac{1}{2} f_1 \wedge b \wedge b \right]_{[\varphi^1][\varphi^2]}\,,\\
	\mathcal{F}_5 + \mathcal{F}_3 \wedge B + \frac{1}{2} F_1 \wedge B \wedge B
	&= f_5 + f_3 \wedge b + \frac{1}{2} f_1 \wedge b \wedge b\\
		& \qquad + \gamma \left[ f_7 + f_5 \wedge b
			+ \frac{1}{2} f_3 \wedge b \wedge b 
			+ \frac{1}{6} f_1 \wedge b \wedge b \wedge b \right]_{[\varphi^1][\varphi^2]}\,,
\end{split}
\end{equation}
while in type IIA string theory we get:
\begin{equation}
\begin{split}
	F_2 &= f_2 + \gamma \left[ f_4 + f_2 \wedge b \right]_{[\varphi^1][\varphi^2]}\,,\\
	\mathcal{F}_4 + F_2 \wedge B 
		&= f_4 + f_2 \wedge b + \gamma \left[ f_6 + f_4 \wedge b
			+ \frac{1}{2} f_2 \wedge b \wedge b \right]_{[\varphi^1][\varphi^2]}\,.
\end{split}
\end{equation}

With a suitable gauge choice, formula~\eqref{RRTsT} can also be recast in terms of R-R potentials:
\begin{equation}\label{RRpotTsT}
	\sum_q C_q \wedge e^{B}
		= \sum_q c_q \wedge e^{b}
		+ \gamma \left[ \sum_q c_q \wedge e^{b} \right]_{[\varphi^1][\varphi^2]}\,,
\end{equation}
which is particularly useful for computations involving the D-brane world-volume action, as we will see in the following.

\section{TsT transformation: world-sheet perspective}\label{s:open}

Let us now study the TsT transformation from the point of view of the world-sheet coordinate fields. In particular, as we have already mentioned, we are interested in studying open string boundary conditions and D-branes in TsT-transformed backgrounds, in order to use them as a tool in the context of the gauge/string duality. The effects of the TsT transformation on the world-sheet have been first studied in~\cite{Frolov:2005dj}, while studies on D-brane transformations can be found for instance in~\cite{Spradlin:2005sv,Imeroni:2006rb}.

We split the world-sheet coordinate fields $X^\mu \equiv \varphi^\mu$ into $\varphi^\mu = (\varphi^1,\varphi^2,\varphi^i)$ where again $\varphi^1$ and $\varphi^2$ are the directions along which the TsT acts with real parameter $\gamma$, and denote the original fields, before the TsT transformation, with a zero subscript.

Using~\eqref{NSNSTsT} and the T-duality rules in the appendix, we derive that under \tst{\varphi^1}{\varphi^2}{\gamma} the fields transform as:
\begin{equation}\label{TsTwv}
\begin{cases}
	\partial_\alpha \varphi^1_{(0)} = \partial_\alpha \varphi^1
		- \gamma B_{2 \mu} \partial_\alpha \varphi^\mu
		- \gamma \eta_{\alpha \beta} \epsilon^{\beta \kappa} G_{2 \mu} \partial_\kappa \varphi^\mu\\
	\partial_\alpha \varphi^2_{(0)} = \partial_\alpha \varphi^2
		+ \gamma B_{1 \mu} \partial_\alpha \varphi^\mu
		+ \gamma \eta_{\alpha \beta} \epsilon^{\beta \kappa} G_{1 \mu} \partial_\kappa \varphi^\mu\\
	\partial_\alpha \varphi^i_{(0)} = \partial_\alpha \varphi^i
\end{cases}\,,
\end{equation}
where $G$ and $B$ are the metric and $B$-field of the TsT-transformed background. For simplicity, we have limited ourselves to the case without world-sheet fermions: the complete expressions can be found in~\cite{Alday:2005ww}.

We wish to use~\eqref{TsTwv} to understand how D-branes transform under TsT. As in the case of a T-duality in flat space, the transformed coordinate fields in~\eqref{TsTwv}, derived using T-duality of the closed string sigma model, are the same ones that are relevant for the open string, and we can therefore use~\eqref{TsTwv} to study open string boundary conditions too. Before the transformation, an open string in the presence of a D-brane extended along the direction $\varphi^{\mu}_{(0)}$ in the original background satisfies (generalized) Neumann boundary conditions:
\begin{equation}\label{Neumann}
	g_{\mu \nu} \partial_\sigma \varphi^\nu_{(0)}
		- \left( b_{\mu \nu} + f_{\mu \nu} \right) \partial_\tau \varphi^\nu_{(0)} = 0\,,
\end{equation}
where $f_{\mu \nu}$ is the gauge field strength on the world-volume. In order to derive the transformation of these boundary conditions we use~\eqref{TsTwv} to compute:
\begin{multline}\label{TsTN}
	g_{\mu \nu} \partial_\sigma \varphi^\nu_{(0)}
		- \left( b_{\mu \nu} + f_{\mu \nu} \right) \partial_\tau \varphi^\nu_{(0)}\\
		= \left[ G_{\mu \nu} + \gamma
			\left( f_{1 \mu} G_{2 \nu} - f_{2 \mu} G_{1 \nu} \right) \right]
			\partial_\sigma \varphi^\nu
			- \left[ B_{\mu \nu} + f_{\mu \nu} 
			+ \gamma \left( f_{1 \mu} B_{2 \nu} - f_{2 \mu} B_{1 \nu} \right) \right]
			\partial_\tau \varphi^\nu\,.
\end{multline}
We stress that $f_{\mu \nu}$ is the field strength on the brane in the undeformed background: it does not match $F$ in the deformed background, that must be computed from~\eqref{TsTN}.

The simplest case one can consider is the one where the initial world-volume field on a D-brane in the undeformed background is zero, $f_{\mu \nu} = 0$\,. In this case, \eqref{TsTN} tells us that Neumann boundary conditions are mapped onto Neumann boundary conditions in the deformed background (with zero gauge field):
\begin{equation}
	G_{\mu \nu} \partial_\sigma \varphi^\nu - B_{\mu \nu} \partial_\tau \varphi^\nu = 0\,,
\end{equation}

An open string in the presence of a brane which is transverse to the direction $\varphi^\mu_{(0)}$ in the undeformed background will instead satisfy Dirichlet boundary conditions:
\begin{equation}
	\partial_\tau \varphi^\mu_{(0)} = 0\,.
\end{equation}
We see from~\eqref{TsTwv} that these boundary conditions are mapped onto the same Dirichlet conditions along $\varphi^\mu$ in the TsT-transformed background, unless the conditions on $\varphi^1_{(0)}$ and $\varphi^2_{(0)}$ are both of Dirichlet type, in which case the resulting set of boundary conditions:
\begin{equation}\label{expand}
\begin{cases}
	G_{2 \nu} \partial_\sigma \varphi^\nu - B_{2 \nu} \partial_\tau \varphi^\nu
		+ \frac{1}{\gamma} \partial_\tau \varphi^1= 0\\
	G_{1 \nu} \partial_\sigma \varphi^\nu - B_{1 \nu} \partial_\tau \varphi^\nu
		- \frac{1}{\gamma} \partial_\tau \varphi^2= 0
\end{cases}
\end{equation}
describes instead Neumann conditions along $\varphi^1$ and $\varphi^2$ with a world-volume gauge field:
\begin{equation}\label{F12}
	F_{12} = \frac{1}{\gamma}
\end{equation}
turned on. This means the D$p$-brane we started from in the undeformed background has turned into a D$(p+2)$-brane wrapped on the two-torus transformed by TsT. Conversely, one can for example see that starting with Neumann boundary conditions along $\varphi^1_{(0)}$ and $\varphi^2_{(0)}$ and gauge field-strength $f_{12} = - \gamma^{-1}$ yields, after the TsT transformation, Dirichlet boundary conditions along $\varphi^1$ and $\varphi^2$, and thus a brane with two fewer longitudinal directions.

It is important to point out that the conditions~\eqref{expand} do not make sense for generic values of the deformation parameter $\gamma$, since the flux~\eqref{F12} along the two-torus must obey a quantization condition. Reinstating the factors of $2 \pi$ in $F$, we see that in order for the undeformed D$p$-brane to expand onto a D$(p+2)$ brane, we should have $\gamma = 1/n$ with $n$ integer. We can generalize this condition by allowing multiple wrappings of the brane on the torus, so that in conclusion the deformation parameter must be rational~\cite{Lunin:2005jy}:
\begin{equation}\label{m/n}
	\gamma = \frac{m}{n}\,.
\end{equation}

What happens to a D-brane transverse to the two-torus if $\gamma$ does not obey~\eqref{m/n}? Using the equations we have derived, we see that the only possibility is for the D-brane to sit at a point where the two-torus $(\varphi^1, \varphi^2)$ shrinks.
%: Figure: TsT transformation of D-branes
\FIGURE{\label{f:Dbrane}
\includegraphics[scale=1]{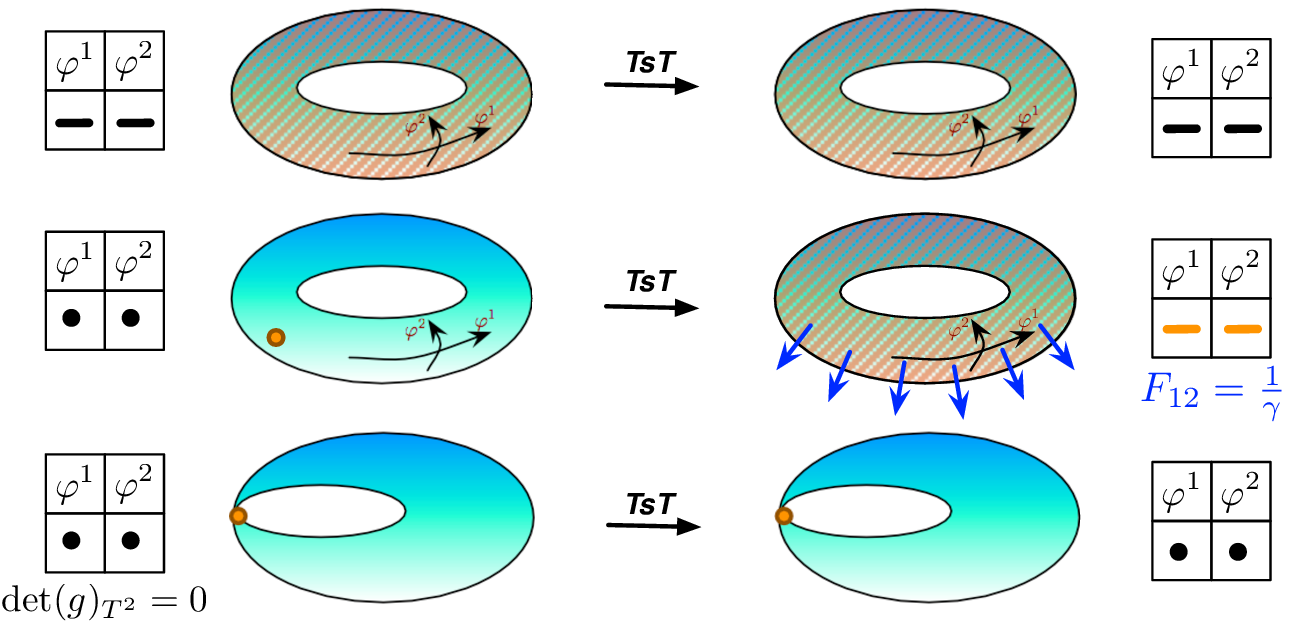}
\caption{TsT transformation of D-branes (with initial $b=0$): a D$p$-brane that extends along the two-torus $(\varphi^1,\varphi^2)$ gets mapped onto the same D$p$-brane in the TsT-transformed background. A D$p$-brane transverse to the two-torus expands (if $\gamma$ is quantized) onto a D$(p+2)$-brane wrapped on the torus with world-volume field strength $F_{12} = 1/\gamma$. For generic $\gamma$, the transformed D-brane must instead sit at a point where the torus shrinks and does not change its dimension after the transformation.}}

Hence the general result is that, under \tst{\varphi^1}{\varphi^2}{\gamma}, a D$p$-brane transverse to the $(\varphi^1, \varphi^2)$ torus will generically be mapped onto the same brane in the TsT-transformed background only if it is placed at the points where the torus is reduced to zero size. If the deformation parameter $\gamma$ is rational, there will moreover be the possibility of having a D$(p+2)$ brane wrapped on the torus with a world-volume flux~\eqref{F12} turned on. This reproduces in much more generality the analysis of~\cite{Imeroni:2006rb}, and we present a summary in figure~\ref{f:Dbrane}. Of course, there will be more complicated cases where one has to compute the resulting boundary conditions by using~\eqref{TsTwv} and~\eqref{TsTN} explicitly.

In the next section, we begin to study, in a very simple case, the behaviour of D-branes and their world-volume actions along the TsT transformation. This sets the stage for the more complicated examples, relevant for the gauge/gravity correspondence, that we will examine in sections~\ref{s:SYM} and~\ref{s:ABJM}.

\section{D-branes and TsT}\label{s:branes}

Before we get to concrete examples of D-brane probes in gauge/gravity duals - it will be our focus in the next sections - let us for now limit ourselves to the simplest possible case in order to analyze general properties of D-brane embeddings in a TsT-transformed background. Let us then start as in~\cite{Lunin:2005jy} (see also~\cite{Spradlin:2005sv}) with the TsT of flat space.

Write the metric of ten dimensional flat Minkowski space as:
\begin{equation}\label{flat}
	ds^2 = dx^2_{1,5} + dr_1^2 + dr_2^2 + r_1^2 (d \varphi^1)^2 + r_2^2 (d \varphi^2)^2\,,
\end{equation}
where we have chosen polar coordinates in $\mathbb{R}^4 \subset \mathbb{R}^{1,9}$. The angles $\varphi^1$ and $\varphi^2$ parameterize two obviously contractible circles. Let us apply a \tst{\varphi^1}{\varphi^2}{\gamma} transformation to this simple background. The formulae in section~\eqref{s:closed} are straightforward in this case and we get:
\begin{equation}\label{TsTflat}
\begin{aligned}
	ds^2 &= dx^2_{1,5} + dr_1^2 + dr_2^2 + \M ( r_1^2 (d \varphi^1)^2 + r_2^2 (d \varphi^2)^2 ) \,,\\
	e^{2 \Phi} &= \M\,,\\
	B &= - \gamma \M r_1^2 r_2^2 d \varphi^1 \wedge d\varphi^2\,,
\end{aligned}
\end{equation}
where
\begin{equation}
	\M^{-1} = 1 + \gamma^2 r_1^2 r_2^2\,.
\end{equation}

We now want to study the fate of a D$p$-brane along the transformation, applying the expertise we got in section~\ref{s:open}. For simplicity, and since it already possesses all the properties we wish to show, we concentrate on a D0-brane. Notice that a D0-brane probe is mainly relevant for the case of the $\beta$-deformation, since the coordinates where the TsT acts are transverse to the world-volume.

Let us then briefly study the action of a D0-brane probe, with world-line coordinate $\sigma^0 = \tau$, in the starting flat space-time. Choosing the static gauge embedding:
\begin{equation}\label{D0emb}
	x^0 = \tau \,,\quad x^i = x^i (x^0)\,,\quad r^i = r^i (x^0)\,,\quad \varphi^i = \varphi^i (x^0)\,,
\end{equation}
The world-volume action~\eqref{Dpwv}
\begin{equation}\label{D0}
	S_{\text{D}0} = - \frac{1}{\gs} \int d\tau \ e^{-\Phi}
		\sqrt{- \det \left(\hat{G}_{ab} + \hat{B}_{ab} \right) }
		 + \frac{1}{\gs} \int \hat{C}_1\,,
\end{equation}
where hats denote pull-backs of the bulk fields onto the one-dimensional world-volume, reduces to:
\begin{equation}\label{D0flat}
	S_{\text{D}0}^{(0)} = - \frac{1}{\gs} \int d\tau \ 
		\sqrt{ 1 - \left( \partial_\tau x^i \partial_\tau x^i
		+ (\partial_\tau r_1)^2 + (\partial_\tau r_2)^2
		+ r_1^2 (\partial_\tau \varphi^1)^2 + r_2^2 (\partial_\tau \varphi^2)^2\right)}\,.
\end{equation}
Neglecting the constant term (that in a more realistic setup, such as the background generated by the supersymmetric D0-brane itself, would be cancelled by the contribution of the R-R part), by expanding the action in derivatives of the fields we see that the geometry of the moduli of the D0-brane theory reproduces, as usual, the geometry of the ambient flat space-time. In particular the $(r_i,\varphi^i)$ part reads the metric of $\mathbb{R}^4$ in polar coordinates.

When we pass to the theory after the deformation, the analysis of the previous section shows that the D0-brane will tend to expand into a D2-brane wrapping the two-torus $(\varphi^1,\varphi^2)$ with a world-volume field $F_{\varphi^1\varphi^2} = 1/\gamma$ turned on, but that this is possible only if $\gamma$ is appropriately quantized to ensure flux quantization on the torus. If $\gamma$ is generic, the D0-brane will not change its dimension along the transformation, but it will be forced to sit at a point where the torus shrinks, as summarized in figure~\ref{f:Dbrane}.

Let us see this explicitly by studying D0 and D2-brane probes in the background~\eqref{TsTflat}. Computing the pull-back of the metric, $B$-field and dilaton onto the world-volume of a D0-brane embedded as in~\eqref{D0emb} we easily get:
\begin{equation}\label{D0flatTsT}
	S_{\text{D}0}^{(\text{TsT})} = - \frac{1}{\gs} \int d\tau
		\sqrt{ \frac{1}{\M} - \frac{\left( \partial_\tau x^i \partial_\tau x^i
		+ (\partial_\tau r_1)^2 + (\partial_\tau r_2)^2 \right)}{\M}
		- \left( r_1^2 (\partial_\tau \varphi^1)^2 + r_2^2 (\partial_\tau \varphi^2)^2 \right) }\,.
\end{equation}
Now we see an angle-dependent potential term that signals the instability of the configuration (and would not be cancelled by the R-R part in more realistic setups). The brane thus has to sit at points where $\M=1$, or equivalently
\begin{equation}
	r_1 = 0 \qquad \text{or} \qquad r_2 = 0\,,
\end{equation}
where the torus along which we performed the TsT transformation shrinks. If we then put $r_1=0$ the action above reduces to:
\begin{equation}
	S_{\text{D}0}^{(\text{TsT})} = - \frac{1}{\gs} \int d\tau
		\sqrt{  1 -  \left(  \partial_\tau x^i \partial_\tau x^i
		+ (\partial_\tau r_2)^2
		+ r_2^2 (\partial_\tau \varphi^2)^2 \right) }\,,
\end{equation}
while if we put $r_2=0$ we get again the same action with the exchange $(r_1,\varphi^1) \leftrightarrow (r_2,\varphi^2)$. Notice that this means that the moduli space of the D0-brane theory does not ``see'' the $\mathbb{R}^4$ part in the $(r_i,\varphi^i)$ sector anymore, but that the moduli space is reduced to $\mathbb{R}^2 \times \mathbb{R}^2$. This is a quite generic feature of moduli spaces on D-brane theories in transformed backgrounds dual to $\beta$-deformations for generic values of the deformation parameter $\gamma$: the moduli space reduces to a sum of complex lines. This is in accordance with gauge theory results when the background in question is dual to some gauge theory, as we will see in the next section.

The moduli space is enlarged when $\gamma$ is quantized, $\gamma = m/n$. In this case we can study a D2-brane probe embedded in~\eqref{TsTflat} as:
\begin{equation}
	x^0 = \sigma^0\,,\quad \varphi^1 = \sigma^1\,,\quad \varphi^2 = \sigma^2\,,\quad
	x^i = x^i (x^0)\,,\quad r^i = r^i (x^0)\,,
\end{equation}
with world-volume gauge field strength:
\begin{equation}
	F_{12} = 1/\gamma\,,\quad
	F_{01} = \partial_0 A_1 (x^0)\,,\quad
	F_{02} = \partial_0 A_2 (x^0)\,.
\end{equation}
$A_1$ and $A_2$ are allowed fluctuations on the brane, exciting periodic Wilson lines on the two-torus, while the constant magnetic flux $F_{12}$ is imposed on us by the transformation rules we have derived. With this embedding, the D2-brane action:
\begin{equation}\label{D2}
	S_{\text{D}2} = - \tau_2 \int d^{3}\sigma\ e^{-\Phi}
		\sqrt{- \det \left(\hat{G}_{ab} + \hat{B}_{ab} + F_{ab}\right) }
		 + \tau_2 \int \left(  \hat{C}_3 + \hat{C}_1 \wedge (\hat{B} + F) \right) \,,
\end{equation}
where $\tau_2 = (4 \pi^2 \gs)^{-1}$, reduces to:
\begin{equation}\label{D2flatTsT}
	S_{\text{D}2}^{(\text{TsT})} = - \frac{\tau_2}{\gamma} \int d^3 \sigma \ 
		\sqrt{ 1 - \left( \partial_\tau x^i \partial_\tau x^i
		+ (\partial_\tau r_1)^2 + (\partial_\tau r_2)^2
		+ \gamma^2
		\left( r_1^2 (\partial_\tau A_2)^2 + r_2^2 (\partial_\tau A_1)^2\right) \right)}\,.
\end{equation}
Notice that the combination $\hat{B}_{12} + F_{12} = \frac{\M - 1}{\gamma} + \frac{1}{\gamma} = \frac{\M}{\gamma}$ is crucial for the cancellations that led us to~\eqref{D2flatTsT} to happen. If we now define:
\begin{equation}\label{ident}
	A_1 = \frac{\phi^2}{\gamma}\,,\qquad A_2 = - \frac{\phi^1}{\gamma}\,,
\end{equation}
and integrate the action over $\sigma^1$ and $\sigma^2$, we get:
\begin{equation}\label{D2flat}
	S_{\text{D}2}^{(\text{TsT})} = - \frac{1}{\gamma} \frac{1}{\gs} \int d\sigma^0 \ 
		\sqrt{ 1 - \left( \partial_\tau x^i \partial_\tau x^i
		+ (\partial_\tau r_1)^2 + (\partial_\tau r_2)^2
		+ r_1^2 (\partial_\tau \phi^1)^2 + r_2^2 (\partial_\tau \phi^2)^2\right)}\,,
\end{equation}
which, apart from an overall $1/\gamma$ factor, precisely coincides with the action~\eqref{D0flat} of a D0-brane in the undeformed flat space-time~\eqref{flat}: $S_{\text{D}2}^{(\text{TsT})} = S_{\text{D}0}^{(0)} / \gamma$. Is the moduli space in the $(r_i,\phi^i)$ sector again the full $\mathbb{R}^4$ here? The identifications~\eqref{ident} show that there is a difference by a factor of $\gamma$ between the periodic Wilson lines on the torus, that have the same period $2 \pi$ as the original coordinates $\varphi^i$, and the new scalars $\phi^i$ of~\eqref{D2flat} that (in the simplest case where we have only one wrapped D2-brane and $\gamma = 1/n$ with $n$ integer) have instead period $2 \pi / n$. We therefore see that the moduli space is reduced from the original flat space to a $\mathbb{Z}_n \times \mathbb{Z}_n$ orbifold. This is also a feature which we will see to be quite general in TsT-transformed backgrounds with rational values $\gamma = m/n$ of the deformation parameter (see also for instance the analysis of~\cite{Butti:2007aq} that is relevant to the TsT of the gravity duals of \Ne{1} quiver gauge theories).

Before we turn to more examples, let us pause for a few considerations. Suppose we want to use a D-brane probe embedded in the background that is generated by the brane itself (or its near-horizon limit). By construction, such a probe computation will yield relevant information on the dual gauge theory, for instance the gauge coupling constant and the theta-angle.

Now suppose that we perform a TsT transformation of the background. The same D-brane probe in the transformed background is supposed to give us information on the corresponding deformed gauge theory. When is the new computation going to be qualitatively different from the one in the undeformed background, giving us new results? In other words, when will the probe be sensitive to the transformation, and when will it instead reproduce the same results as in the original undeformed background?

The results of this section and of the previous one hint to the following general perspective: if the D-brane probe that was used in the undeformed background was wrapped along one or more directions of the torus $(\varphi^1, \varphi^2)$, then the TsT transformation will leave the brane untouched and the result of the evaluation of the action of the probe in the deformed background is likely to be the same. On the contrary, the situation becomes more interesting when the D-brane before the transformation is transverse to the two-torus. After TsT, the brane will either be stuck at a point where the torus shrinks, or will expand onto a higher dimensional brane wrapped on the torus. This means that the results for the gauge theory will be also quite different from the ones obtained in the undeformed background!

In this respect, it is clear that the most interesting situation will be realized in the case of $\beta$-deformations~\eqref{starbeta}, rather than in non-commutative or dipole deformations, because in such case our D-brane probe will be transverse to the two-torus where the transformation acts, and will thus exhibit the more varied behaviour we have just described. Of course, one can envision different gauge theory quantities that will be read on probes that are not the D-branes generating the background but, since in this paper we will limit ourselves to this kind of probes, in the following examples we will devote much more attention to $\beta$-deformations, while just presenting the deformed backgrounds dual to non-commutative and dipole deformations.

\section{Example 1: \Ne{4} Super Yang-Mills}\label{s:SYM}

After having studied D-brane probes in the TsT deformation of flat space, we pass to our first simple example in the context of the gauge/string duality. We will consider deformations of \Ne{4} Super Yang-Mills theory in four dimensions and of its \ads{} gravity dual. This is mostly review material and the backgrounds we are going to present are already known, but we will rederive them in a unified way by making use of the formulae we presented in section~\ref{s:closed}, then we will give some examples on how TsT transformed D-brane probes can be used to make contact between the two sides of the correspondence.

Let us start with the undeformed gauge theory. The vector multiplet of \Ne{4} SYM can be decomposed in such a way that, in \Ne{1} notation, it contains three chiral superfields $\Phi_i$ subject to the superpotential:
\begin{equation}\label{N=4W}
	W = \tr\ (\Phi_1 \Phi_2 \Phi_3 - \Phi_1 \Phi_3 \Phi_2)\,,
\end{equation}
that we will keep in mind, in order to deform it later.

The \ads{} solution of type IIB supergravity, which is dual to \Ne{4} SYM, reads:
\begin{equation}\label{ads5s5}
\begin{aligned}
	ds^2 &= R^2 \left( ds^2_{AdS_5} + ds^2_{S^5} \right) \,, \\
	e^{\Phi} &= 1\,, \\
	F_5 &= -4 R^4 \left( \omega_{AdS_5} + \omega_{S^5} \right)\,,
\end{aligned}
\end{equation}
The common radius of $AdS_5$ and $S^5$ is given by $R^4 = 4 \pi \gs N$ (in units where $\ls = 1$). For the unit radius $AdS_5$ space we will use Poincar\'e coordinates, with metric and volume form given by:
\begin{equation}\label{adspoinc}
\begin{aligned}
	ds^2_{AdS_5} &= r^2 dx^2_{1,3}  + \frac{dr^2}{r^2}\,, \\
	\omega_{AdS_5} &= - r^3 dx^0 \wedge dx^1 \wedge dx^2 \wedge dx^3 \wedge dr\,,
\end{aligned}
\end{equation}
where $dx^2_{1,3}$ is the flat Minkowski metric in four dimensions. The metric and the corresponding volume form we will use for the five-sphere are given by:
\begin{equation}\label{S5}
\begin{aligned}
	ds^2_{S^5} &= \sum_{i=1}^{3} \left(d \mu_i^2+ \mu_i^2 d\phi_i^2\right)\,, 
		\qquad \sum_i \mu_i^2 = 1\,, \\
	\omega_{S^5} &= \cos \alpha \sin^3 \alpha \sin \theta \cos \theta d \alpha \wedge d \theta
		\wedge d\phi^1 \wedge d\phi^2 \wedge d\phi^3\,,
\end{aligned}
\end{equation}
where we have chosen the parameterization:
\begin{equation}
	\mu_1 = \cos \alpha\,,\qquad
	\mu_2 = \sin \alpha \cos \theta\,,\qquad
	\mu_3 = \sin \alpha \sin \theta\,.
\end{equation}

We will now study deformations of the \ads{} solution. As we said at the end of section~\ref{s:branes}, the D-brane probes we will use to elucidate properties of the deformed gauge theories are particularly interesting in the case of the $\beta$-deformation. We will therefore start  in subsection~\ref{s:SYMbeta} with the $\beta$-deformation, before briefly presenting other cases for completeness.

\subsection{$\beta$-deformation}\label{s:SYMbeta}

%: Table: N=4 SYM charges
\TABLE{
\begin{tabular}{|c|c|c|c|}
\hline
& $\Phi_1$ & $\Phi_2$ & $\Phi_3$ \\
\hline
$U(1)_{\varphi_1}$ & $0$ & $+1$ & $-1$ \\
\hline
$U(1)_{\varphi_2}$ & $-1$ & $+1$ & $0$ \\
\hline
\end{tabular}
\caption{$U(1) \times U(1)$ charges of the chiral fields of \Ne{4} Super Yang-Mills.\label{t:SYM}}}
We consider the the well-known $\beta$-deformation of \Ne{4} Super Yang-Mills~\cite{Leigh:1995ep}. The deformation is obtained by starting from the superpotential~\eqref{N=4W} and modifying it as follows:
\begin{equation}\label{N=4Wbeta}
	W \to W_{\gamma} = \tr\ (e^{i \pi \gamma} \Phi_1 \Phi_2 \Phi_3
		- e^{- i \pi \gamma} \Phi_1 \Phi_3 \Phi_2)
\end{equation}
(remember we limit ourselves to the case where the deformation parameter is real). The deformed superpotential~\eqref{N=4Wbeta} can also be seen as being obtained by replacing the ordinary products in the lagrangian of \Ne{4} SYM by the star product~\eqref{starbeta}, where the charges of the three chiral superfields $\Phi_i$ under the two $U(1)$ groups are the ones listed in table~\ref{t:SYM}. One can show that the $\beta$-deformed theory is an exactly marginal deformation of \Ne{4} SYM preserving \Ne{1} superconformal theory~\cite{Leigh:1995ep}.

From the point of view of geometry, it is clear that in order to obtain the gravity dual we have to perform a TsT transformation of \ads{} along two directions $\varphi_1$ and $\varphi_2$ corresponding to the two $U(1)$ factors in table~\ref{t:SYM}~\cite{Lunin:2005jy}. This can be achieved through the following change of coordinates on the $S^5$:
\begin{equation}\label{varphi}
	\phi_1 = \varphi_3 - \varphi_2\,,\qquad
	\phi_2 = \varphi_3 + \varphi_1 + \varphi_2\,,\qquad
	\phi_3 = \varphi_3 - \varphi_1\,,
\end{equation}
so that we can rewrite the metric~\eqref{S5} as:
\begin{multline}\label{S52}
	ds^2_{S^5} = d\alpha^2 + \sin^2\alpha d\theta^2 + \cos^2 \alpha ( d \varphi_3 - d \varphi_2)^2\\
		+ \sin^2 \alpha \cos^2 \theta ( d \varphi_3 + d \varphi_1 + d \varphi_2 )^2
		+ \sin^2 \alpha \sin^2 \theta ( d \varphi_3 - d \varphi_1 )^2\,.
\end{multline}

We can now perform the \tst{\varphi_1}{\varphi_2}{\gamma} transformation on the solution~\eqref{ads5s5} with the $S^5$ metric in the form~\eqref{S52}. The resulting transformed background reads:
\begin{equation}\label{N=4beta}
\begin{aligned}
	ds^2 &= R^2 \left( ds^2_{AdS_5} + ds^2_{\widetilde{S^5}} \right) \,, \\
	ds^2_{\widetilde{S^5}} &= d\alpha^2 + \sin^2\alpha d\theta^2
		+ \M \big(\cos^2 \alpha ( d \varphi_3 - d \varphi_2)^2
		+ \sin^2 \alpha \cos^2 \theta ( d \varphi_3 + d \varphi_1 + d \varphi_2 )^2\\
		& \qquad + \sin^2 \alpha \sin^2 \theta ( d \varphi_3 - d \varphi_1 )^2 
		+ 9 \hgamma^2 \cos^2 \alpha \sin^4 \alpha \cos^2 \theta \sin^2 \theta d\varphi_3^2 \big)\,,\\
	e^{2 \Phi} &= \M\,, \\
	B &= - \hgamma R^2 \M \sin^2 \alpha 
		\Big[ ( \cos^2 \alpha + \sin^2 \alpha \cos^2 \theta \sin^2 \theta)
			d \varphi_1 \wedge d \varphi_2 \\
		&\qquad + ( \sin^2\alpha \cos^2\theta \sin^2\theta+ \cos^2\alpha \sin^2 \theta
			- 2 \cos^2 \alpha \cos^2 \theta ) d \varphi_2 \wedge d \varphi_3 \\
		&\qquad+ ( \cos^2\alpha \cos^2\theta + \sin^2\alpha \sin^2 \theta
			- 2 \sin^2 \alpha \cos^2 \theta \sin^2\theta) d \varphi_3 \wedge d \varphi_1 \Big] \,,\\
	C_2 &=  3 \hgamma R^4 \sin^4 \alpha \sin \theta \cos \theta d\theta \wedge d\varphi_3\,,\\
	F_5 &= -4 R^4 \left( \omega_{AdS_5} + \M \omega_{S^5} \right)\,,
\end{aligned}
\end{equation}
where we have defined $\hgamma = R^2 \gamma$ and:
\begin{equation}\label{Mbeta}
	\M^{-1} = 1 + \hgamma^2 \sin^2 \alpha ( \cos^2 \alpha + \sin^2 \alpha \cos^2 \theta \sin^2 \theta)\,.
\end{equation}
This is the Lunin-Maldacena solution~\cite{Lunin:2005jy}. The $AdS_5$ factor is preserved by the transformation, which is the gravity dual counterpart of the gauge theory statement that the $\beta$-deformation is an exactly marginal deformation of \Ne{4} SYM.

We can rewrite the transformed background~\eqref{N=4beta} in a more symmetric way by reverting to the original coordinates $\phi_i$ in~\eqref{varphi}. The change of coordinates yields:
\begin{equation}\label{LM}
\begin{aligned}
	ds^2 &= R^2  \left[  ds^2_{AdS_5} + \sum_{i=1}^3 \left(d \mu_i^2+ \M \mu_i^2 d\phi_i^2\right)
		+ \hgamma^2 \M \mu_1^2 \mu_2^2 \mu_3^2 \big( \sum_i d\phi_i \big)^2 
		\right]\,,\\
	e^{2\Phi} &= \M \,,\\
	B &=  - \hgamma R^2 \M  \left(\mu_1^2 \mu_2^2 d\phi_1 \wedge d\phi_2
		+ \mu_2^2 \mu_3^2 d\phi_2 \wedge d\phi_3
		+ \mu_3^2 \mu_1^2 d\phi_3 \wedge d\phi_1\right)\,,\\
	C_2 &=  \hgamma R^2 \sin^4 \alpha \sin \theta \cos \theta
		d\theta \wedge (d\phi_1 + d\phi_2 + d\phi_3)\,,\\
	F_5 &= -4 R^4 \left( \omega_{AdS_5} + \M \omega_{S^5} \right)\,,
\end{aligned}
\end{equation}
where:
\begin{equation}\label{MLM}
	\M^{-1} = 1 + \hat{\gamma}^2 (\mu_1^2 \mu_2^2 + \mu_2^2 \mu_3^2 + \mu_3^2 \mu_1^2)\,.
\end{equation}

We are now ready to study some features of the duality for the deformed gauge theory: we will study the vacuum structure as determined by the study of static D-brane probes. The origin of the undeformed \ads{} background can be traced to the backreaction of a stack of D3-branes, so embedding a D3-brane probe in the solution will ``read'' the moduli space of the theory, that is just a symmetrized product of $N$ copies of $\mathbb{C}^3$. What happens after the deformation? As in the simple flat space case of section~\ref{s:branes}, we have to study D3 and D5-brane probes in the background~\eqref{LM}.

Let us start with a D3-brane. We choose the static gauge $\sigma^a = x^a$ ($a=0,1,2,3$) and for now we keep $r$ and the coordinates on the deformed sphere fixed. The D3-brane action, once we also choose the world-volume gauge field $F = 0$, is:
\begin{equation}\label{D3}
	S_{\text{D}3} = - \tau_3 \int d^{4}\sigma\ e^{-\Phi}
		\sqrt{- \det \left(\hat{G}_{ab} + \hat{B}_{ab} \right) }
		 + \tau_3 \int \left( \hat{C}_4 + \hat{C}_2 \wedge \hat{B}  \right)\,,
\end{equation}
where $\tau_3 = (8 \pi^3 \gs)^{-1}$. We immediately see that the Wess-Zumino part reduces to the one of a D3-brane in the undeformed \ads{} background, $\tau_3 \int \hat{C}_4^{(0)} = R^4 r^4$, since one can verify that $\hat{C}_4 + \hat{C}_2 \wedge \hat{B} =  \hat{C}_4^{(0)}$, as encoded in the general formula~\eqref{RRpotTsT}. The DBI part instead contains a $\gamma$-dependent factor, so that the result is:
\begin{equation}
	S_{\text{D}3}^{\text{TsT}} = - \frac{N}{2 \pi^2} \int d^{4}\sigma\ r^4 ( \M^{-1/2} - 1 )\,.
\end{equation}
Looking at equation~\eqref{MLM} we see that there is an angle-dependent potential that we need to cancel to make the probe supersymmetric and stable. This is a concrete example of what we have seen in general in section~\ref{s:open}: the D3-brane has to sit at a point where the torus touched by the TsT transformation shrinks, so that $\M = 1$. There are three possibilities for this to happen:
\begin{equation}\label{3cases}
\begin{aligned}
	&\text{(i)}& \mu_1&=1\,,& & \mu_2=\mu_3=0 &  &\left(\alpha=0\right)\,,\\
	&\text{(ii)}& \mu_2&=1\,,& & \mu_3=\mu_1=0& &\left(\alpha=\tfrac{\pi}{2}\,,\:\: \theta=0\right)\,,\\
	&\text{(iii)}& \mu_3&=1\,,& & \mu_1=\mu_2=0&
		&\left(\alpha=\tfrac{\pi}{2}\,,\:\: \theta=\tfrac{\pi}{2}\right)\,.
\end{aligned}
\end{equation}
Let us for instance choose case (i) above, where $\alpha = 0$. Repeating the probe computation by allowing the transverse coordinates $r$, $\theta$ and $\phi_i$ to depend on the world-volume coordinates $\sigma^a$ and expanding the result for slowly varying fields we get:
\begin{equation}\label{R2}
	S_{\text{D}3}^{\text{TsT}} = - \frac{N}{2 \pi^2} \int d^{4}\sigma\  \frac{1}{2} \left( ( \partial_a r )^2
		+ r^2 ( \partial_a \phi_1 )^2 \right)\,.
\end{equation}
The moduli space of the theory on this probe is simply $\mathbb{R}^2$ written in polar coordinates $(r, \phi_1)$. We can repeat the computation for the cases (ii) and (iii) in~\eqref{3cases} to find two more copies of $\mathbb{R}^2$, with polar angles $\phi_2$ and $\phi_3$. This translates in the gauge theory as the statement that, for generic deformation parameter $\gamma$, the abelian moduli space is reduced from $\mathbb{C}^3$ for \Ne{4} SYM to three copies of $\mathbb{C}$ for the $\beta$-deformed theory. These branches of the moduli space are the solution of the three F-term equations derived from the superpotential~\eqref{N=4Wbeta}:
\begin{equation}
	\Phi_2 \Phi_3 + b \Phi_3 \Phi_2 = 0 \,,\quad
	\Phi_3 \Phi_1 + b \Phi_1 \Phi_3 = 0 \,,\quad
	\Phi_1 \Phi_2 + b \Phi_2 \Phi_1 = 0 \,,
\end{equation}
where $b = e^{-i 2 \pi \gamma}$ and $\gamma$ is generic.

However, new branches of vacua open up when $\gamma$ is rational, $\gamma = m/n$ \cite{Berenstein:2000hy}. In this case, we know from our previous analysis that we can consider D5-brane probes, that extend along the torus $( \varphi_1 , \varphi_2 )$  where the TsT transformation has been performed. We then consider the following D5-brane embedding:
\begin{equation}
\begin{gathered}
	x^a = \sigma^a\ (a=0,1,2,3)\,,\quad
	\varphi_1 = \sigma^4\,,\quad
	\varphi_2 = \sigma^5\,,\\
	r = r (x^a)\,, \quad
	\alpha = \alpha (x^a)\,, \quad
	\theta = \theta (x^a)\,,  \quad
	\varphi_3 = \varphi_3 (x^a)\,,
\end{gathered}
\end{equation}
and world-volume field strength:
\begin{equation}
	F_{45} = 1 / \gamma\,,\qquad
	F_{a4} = \partial_a A_4 (x^a) \,,\qquad
	F_{a5} = \partial_a A_5 (x^a)\,.
\end{equation}
The D5-brane action is given by:
\begin{multline}\label{D5}
	S_{\text{D}5} = - \tau_5 \int d^{6}\sigma\ e^{-\Phi}
		\sqrt{- \det \left(\hat{G}_{ab} + \hat{B}_{ab} + F_{ab}\right) }\\
		 + \tau_5 \int \left( \hat{C}_6 + \hat{C}_4 \wedge ( \hat{B} + F )
		  + \frac{1}{2} \hat{C}_2 \wedge ( \hat{B} + F ) \wedge ( \hat{B} + F ) \right) \,.
\end{multline}
We start from the Wess-Zumino part, that in our background~\eqref{LM} where $B \wedge B = 0$ can be written as:
\begin{equation}
	S_{\text{D}p}^{\text{WZ}} = \tau_5 \int \left[ ( \hat{C}_6 + \hat{C}_4 \wedge \hat{B} )
		  + ( \hat{C}_4 + \hat{C}_2 \wedge \hat{B} ) \wedge F \right]\,.
\end{equation}
Now, either by direct computation or more easily with the aid of the general expression~\eqref{RRpotTsT}, we see that we can choose a gauge where $C_6 + C_4 \wedge B$ vanishes, while $C_4 + C_2 \wedge B$ reduces once again to the four-form potential $C_4^{(0)}$ of the undeformed \ads{} solution. Hence the Wess-Zumino part of the action, integrated along $\varphi_1$ and $\varphi_2$, reduces to the one of a D3-brane in the undeformed background divided by a factor of $\gamma$.

Now we can compute the determinant of the Dirac-Born-Infeld factor and write down the full action. We integrate the result along $\sigma^4$ and $\sigma^5$ to find:
\begin{equation}\label{D5TsTfinal}
\begin{split}
	S_{\text{D}5}^{\text{TsT}} &= - \frac{1}{\gamma} \frac{N}{2 \pi^2} \int d^{4}\sigma\ r^4 \bigg[
		 \Big( 1 + \frac{1}{r^4} \Big( (\partial_a r)^2
		+ r^2 \Big( (\partial_a \alpha)^2
		+ \sin^2 \alpha (\partial_a \theta)^2\\
		&\qquad + \cos^2 \alpha ( \partial_a \varphi_3 + \gamma \partial_a A_4)^2
		+ \sin^2 \alpha \cos^2\theta ( \partial_a \varphi_3
			+  \gamma \partial_a A_5 - \gamma \partial_a A_4)^2\\
		&\qquad + \sin^2 \alpha \sin^2\theta (\partial_a \varphi_3 
			- \gamma \partial_a A_5)^2
		\Big) \Big) \Big)^{1/2}-1\bigg] \,.
\end{split}
\end{equation}
If we now introduce new scalar fields:
\begin{equation}\label{varphiA}
	\varphi_1 = \gamma A_5 \,,\qquad \varphi_2 = - \gamma A_4\,,
\end{equation}
we recognize in the expansion of the square root in~\eqref{D5TsTfinal} the flat space metric expressed as a cone over the undeformed five-sphere~\eqref{S52}. We should not forget however that the identifications~\eqref{varphiA} have modified the periodicities of two angles on the sphere by a factor of the denominator $n$ of the rational deformation parameter $\gamma$. The resulting abelian moduli space in these additional branches is then a $\mathbb{Z}_n \times \mathbb{Z}_n$ orbifold of the undeformed moduli space $\mathbb{C}^3$. The full moduli space is then obtained by means of a symmetrized product as in the undeformed case. This is in accordance with the known properties of the $\beta$-deformed gauge theory~\cite{Berenstein:2000hy,Berenstein:2000ux,Dorey:2004xm,Benini:2004nn}.

One may consider other probes in the Lunin-Maldacena background~\eqref{LM}, and in particular giant gravitons have been shown to be relevant for the study of the theory on $S^3$~\cite{Pirrone:2006iq,Imeroni:2006rb,Pirrone:2008av}. In that case, an explicit map was constructed between D5-brane dual giant gravitons, wrapped on the two-torus $(\varphi_1,\varphi_2)$, that are stable when $\gamma$ is rational, and rotating expectation values in the additional branches of the gauge theory~\cite{Imeroni:2006rb}. Other interesting objects are D7-branes, whose world-volume fluctuations, studied exhaustively in~\cite{Penati:2007vj}, correspond to mesonic excitations of the flavored $\beta$-deformed theory.

\subsection{3-parameter deformation}\label{s:SYMbeta3}

In the previous subsection, we have considered a marginal deformation of \Ne{4} SYM preserving \Ne{1} supersymmetry. If we allow for a complete breaking of supersymmetry, though, we can find a more general 3-parameter family of deformations, first derived in~\cite{Frolov:2005dj}. In fact, we can form three distinct two-tori from the angles $\phi_i$ on the five-sphere~\eqref{S5} and perform successive TsT transformations on all of them.

Specifically, we perform a \tst{\phi_1}{\phi_2}{\gamma_3} transformation, followed by \tst{\phi_2}{\phi_3}{\gamma_1} and finally by  \tst{\phi_3}{\phi_1}{\gamma_2}. The resulting type IIB supergravity solution, that should be dual to a non-supersymmetric but marginal deformation of \Ne{4} SYM, reads:
\begin{equation}\label{Frolov}
\begin{aligned}
	ds^2 &= R^2 \left[  ds^2_{AdS_5} + \sum_{i=1}^3 \left(d \mu_i^2+ \M \mu_i^2 d\phi_i^2\right)
		+ \M \mu_1^2 \mu_2^2 \mu_3^2 \big( \sum_i \hgamma_i d\phi_i \big)^2 
		\right]\,,\\
	e^{2\Phi} &= \M\,,\\
	B &= - R^2 \M (\hgamma_3 \mu_1^2 \mu_2^2 d\phi_1 \wedge d\phi_2
		+ \hgamma_1 \mu_2^2 \mu_3^2 d\phi_2 \wedge d\phi_3
		+ \hgamma_2 \mu_3^2 \mu_1^2 d\phi_3 \wedge d\phi_1)\,,\\
	C_2 &= R^2 \sin^4 \alpha \sin \theta \cos \theta d\theta
		\wedge (\hgamma_1 d\phi_1 +\hgamma_2  d\phi_2 +\hgamma_3  d\phi_3)\,,\\
	F_5 &= -4 R^4 \left( \omega_{AdS_5} + \M \omega_{S^5} \right)\,,
\end{aligned}
\end{equation}
where now $\M$ is given by:
\begin{equation}
	\M^{-1} = 1 + (\hgamma_3^2 \mu_1^2 \mu_2^2
		+\hgamma_1^2  \mu_2^2 \mu_3^2 +\hgamma_2^2  \mu_3^2 \mu_1^2)\,.
\end{equation}
The supersymmetric solution~\eqref{LM} of~\cite{Lunin:2005jy} is recovered from~\eqref{Frolov} by setting all deformation parameters equal, $\hgamma_1=\hgamma_2=\hgamma_3=\hgamma$. We will not repeat here probe computations analogous to the ones we have done for the supersymmetric $\beta$-deformation, but also in this case there are D5-branes indicating the existence of additional branches of the theory. For instance D5-brane dual giant gravitons were considered in~\cite{Imeroni:2006rb}.

\subsection{Non-commutative deformation}\label{s:SYMnc}

In this and in the next subsection, we wish to rederive some known backgrounds in order to show how our general framework correctly accounts for the duals to non-commutative and dipole theories. Let us start with the dual to non-commutative \Ne{4} Super Yang-Mills. Suppose we want to put the gauge theory on a non-commutative two-torus:
\begin{equation}\label{ncSYM}
	[ x^1 , x^2 ] = i \theta^{12}\,.
\end{equation}
The discussion in section~\ref{s:gauge} then instructs us to compactify the coordinates $x^1$ and $x^2$ of the $AdS_5$ space~\eqref{adspoinc}, and to perform a TsT transformation \tst{x^1}{x^2}{\gamma} such that the parameter $\gamma$ is related to the non-commutativity parameter in~\eqref{ncSYM} by:
\begin{equation}
	\theta^{12} = - 2 \pi \gamma\,.
\end{equation}

We then apply the formulae in section~\ref{s:closed}. The resulting TsT-transformed solution is given by:
\begin{equation}\label{N=4nc}
\begin{aligned}
	ds^2 &= R^2 \left( r^2 (- (dx^0)^2 + (dx^3)^2 
		+ \M ( (dx^1)^2 + (dx^2)^2)
		+ \frac{dr^2}{r^2} + ds^2_{S^5} \right) \,, \\
	e^{2 \Phi} &= \M\,, \\
	B &= - \hgamma \M R^2 r^4 dx^1 \wedge dx^2\,,\\
	C_2 &= \hgamma R^2 r^4 dx^0 \wedge dx^3\,,\\
	F_5 &= -4 R^4 \left( \M \omega_{AdS_5} + \omega_{S^5} \right)\,,
\end{aligned}
\end{equation}
where we have defined $\hgamma = R^2 \gamma$ and
\begin{equation}
	\M^{-1} = 1 + \hgamma^2 r^4\,.
\end{equation}
The solution~\eqref{N=4nc} is the same solution obtained in~ \cite{Hashimoto:1999ut,Maldacena:1999mh} as the dual of non-commutative Super Yang-Mills, but our TsT technology has allowed us to get it easily in a single step. We will not analyze the theory in more detail here because, as we noticed in section~\ref{s:branes}, the D-brane probes we are interested in do not give any particular new insight in the case where the TsT transformation acts completely along the world-volume. So for additional studies of non-commutative SYM and its gravity dual we refer the reader to the vast literature.

\subsection{Dipole deformation}\label{s:SYMdipole}

The gravity dual of the dipole theory is obtained by performing a TsT transformation along a direction longitudinal to the D-brane supporting the gauge theory and along a $U(1)$ transverse isometry. As an example, we perform the transformation \tst{x^3}{\phi_1}{\gamma}, where $x^3$ and $\phi_1$ are respectively a coordinate of $AdS_5$ and a coordinate of $S^5$ in~\eqref{ads5s5}.

The resulting solution, dual to a dipole deformation of \Ne{4} SYM, reads:
\begin{equation}\label{N=4dipole}
\begin{aligned}
	ds^2 &= R^2 \left( r^2 ( dx^2_{1,2}
		+ \M (dx^3)^2 )
		+ \frac{dr^2}{r^2} + d \mu_1^2+ \M \mu_1^2 d\phi_1^2
		+ \sum_{i=2}^{3} \left(d \mu_i^2+ \mu_i^2 d\phi_i^2\right) \right) \\
	e^{\Phi} &= \M\,, \\
	B &= - \hgamma \M R^2 r^2 \mu_1^2\,,\\
	F_5 &= -4 R^4 \left( \omega_{AdS_5} + \omega_{S^5} \right)\,,
\end{aligned}
\end{equation}
where we have again defined $\hgamma = R^2 \gamma$ and:
\begin{equation}
	\M^{-1} = 1 + \hgamma^2 r^2 \mu_1^2\,. 
\end{equation}
Notice that the R-R part of the solution is untouched by this transformation since $F_5$ has no components mixing the $AdS$ and sphere parts.

This is of course not the most symmetric dipole deformation we could have derived. In an analogous way, one can write down different dipole deformations by performing a  \tst{x^3}{\varphi}{\gamma} transformation, $\varphi$ being a linear combination of the angles $\phi_i$.

\section{Example 2: \Ne{6} Chern-Simons-matter theory}\label{s:ABJM}

We now turn to our second example.  Motivated by the work of Bagger, Lambert~\cite{Bagger:2006sk,Bagger:2007jr,Bagger:2007vi} and Gustavsson~\cite{Gustavsson:2007vu,Gustavsson:2008dy}, which opened the way to the study of three-dimensional superconformal theories and their relation with M2-branes, Aharony, Bergman, Jafferis and Maldacena (ABJM) constructed an \Ne{6} superconformal Chern-Simons-matter theory and conjectured it to be the theory living on $N$ M2-branes probing a $\cZ{4}{k}$ singularity in M-theory~\cite{Aharony:2008ug}.

Their work was the starting point of a long series of developments. Some studies of the gravity dual of the ABJM theory, and in particular of its deformations, that are akin to the spirit of our work in this section can be found in~\cite{Berman:2007tf,Berman:2008be,Benna:2008zy,Nishioka:2008gz,Armoni:2008kr,Ahn:2008gda,Grignani:2008is,Hosomichi:2008jb,Grignani:2008te,Gomis:2008vc,Hashimoto:2008iv,Astolfi:2008ji,Aharony:2008gk,Ooguri:2008dk,Jafferis:2008qz,Martelli:2008rt,Martelli:2008si}.

The gauge group of the ABJM theory is $U(N) \times U(N)$ and the two Chern-Simons factors have opposite levels $k$ and $-k$. The matter part comprises two chiral superfield $A_1$ and $A_2$ transforming in the  bifundamental $(\mathbf{N},\mathbf{\bar{N}})$ representation, and two chiral superfield $B_1$ and $B_2$ transforming in the anti-bifundamental $(\mathbf{\bar{N}},\mathbf{N})$ representation. There is a quartic superpotential given by:
\begin{equation}\label{M2W}
	W = \frac{4 \pi}{k} \tr \left( A_1 B_1 A_2 B_2 - A_1 B_2 A_2 B_1 \right)\,,
\end{equation}
which can also be seen as the three-dimensional version of the four-dimensional superpotential of the theory living on D3-branes at a conifold singularity~\cite{Klebanov:1998hh}. As we said, this is a superconformal theory preserving \Ne{6} supersymmetry in three dimensions

The theory is weakly coupled when $k \gg N$. The gravity dual of the \Ne{6} Chern-Simons-matter theory is, as usual, valid in the 't Hooft limit, which in this case is:
\begin{equation}
	N\,, k \to \infty\,,\qquad
	\lambda = \frac{N}{k}\ \ \text{fixed}.
\end{equation}
It was shown in~\cite{Aharony:2008ug} that the appropriate dual gravitational description depends on the range of the parameters $N$ and $k$. In particular, for $N \gg k^5$ the gravity dual is the \adsk{} solution of eleven dimensional supergravity, while for $k \ll N \ll k^5$ the appropriate description is in term of the $AdS_4 \times \cp$ solution of type IIA supergravity. We now review the ten and eleven-dimensional solutions, in order to deform them via TsT transformations in the following subsections.

The $AdS_4 \times \cp$ solution of type IIA supergravity can be written in terms of $k$ and $R = 32 \pi^2 k N$ as (see for instance~\cite{Nilsson:1984bj,Watamura:1983hj,Nishioka:2008gz}):
\begin{equation}\label{ads4cp3}
\begin{aligned}
	ds^2_{\text{IIA}} &= \frac{R^3}{k} \left( \frac{1}{4} ds^2_{AdS_4} + ds^2_{\cp} \right)\,,\\
	ds^2_{\cp} &= d\xi^2 + \cos^2 \xi \sin^2 \xi
		\left( d\psi + \frac{1}{2} \cos \theta_1 d\varphi_1 - \frac{1}{2} \cos \theta_2 d\varphi_2 \right)^2\\
		&\qquad + \frac{1}{4} \cos^2 \xi (d\theta_1^2 + \sin^2 \theta_1 d\varphi_1^2 )
		 + \frac{1}{4} \sin^2 \xi (d\theta_2^2 + \sin^2 \theta_2 d\varphi_2^2 )\,,\\
	e^{2 \Phi} &= \frac{R^3}{k^3}\,,\\
	C_1 &= \frac{k}{2} \left( ( \cos^2 \xi - \sin^2 \xi ) d\psi
		+ \cos^2 \xi \cos \theta_1 d \varphi_1
		+ \sin^2 \xi \cos \theta_2 d \varphi_2 \right)\,,\\
	F_2 &= k  \Big( - \cos \xi \sin \xi d\xi
		\wedge (2 d\psi + \cos \theta_1 d \varphi_1 - \cos \theta_2 d \varphi_2)\\
		&\qquad -\frac{1}{2} \cos^2\xi \sin\theta_1 d\theta_1 \wedge d\varphi_1
		-\frac{1}{2} \sin^2\xi \sin\theta_2 d\theta_2 \wedge d\varphi_2\Big)\,,\\
	F_4 &= - \frac{3 R^3}{8} \omega_{AdS_4}\,.
\end{aligned}
\end{equation}
For $AdS_4$ we will use the Poincar\'e patch, so:
\begin{equation}\label{ads4poinc}
\begin{aligned}
	ds^2_{AdS_4} &= r^2 dx^2_{1,2}  + \frac{dr^2}{r^2}\,, \\
	\omega_{AdS_4} &= r^2 dx^0 \wedge dx^1 \wedge dx^2 \wedge dr\,,
\end{aligned}
\end{equation}
where $dx^2_{1,2}$ is the Minkowski metric in three dimensions.

The \adsk{} solution of eleven dimensional supergravity is given by:
\begin{equation}\label{ads4s7k}
\begin{aligned}
	ds^2_{11} &= \frac{R^2}{4} ds^2_{AdS_4} + R^2 ds^2_{\sk}\,,\\
	ds^2_{\sk} &= ds^2_{\cp} + \frac{1}{k^2} ( dy + C_1 )^2\,,\\
	G_4 &= - \frac{3 R^3}{8} \omega_{AdS_4}\,,
\end{aligned}
\end{equation}
where $ds^2_{\cp}$ and $C_1$ are quantities given in the type IIA solution~\eqref{ads4cp3}.

It will be useful for what follows to recall the relation between the coordinates we have used on \sk{} and the supergravity fields that correspond to the bifundamentals $A_i$ and $B_i$ in~\eqref{M2W}. The embedding equations can be written as~\cite{Nishioka:2008gz}:
\begin{equation}\label{s7emb}
\begin{aligned}
	A_1 &= \cos \xi \cos \tfrac{\theta_1}{2}
		e^{i \left( \frac{y}{k} + \frac{\psi}{2} + \frac{\varphi_1}{2}\right)}\,, &
	A_2 &= \cos \xi \sin \tfrac{\theta_1}{2}
		e^{i \left( \frac{y}{k} + \frac{\psi}{2} - \frac{\varphi_1}{2}\right)}\,, \\
	B_1 &= \sin \xi \cos \tfrac{\theta_2}{2}
		e^{- i \left( \frac{y}{k} - \frac{\psi}{2} + \frac{\varphi_2}{2}\right)}\,, &
	B_2 &= \sin \xi \sin \tfrac{\theta_2}{2}
		e^{- i \left( \frac{y}{k} - \frac{\psi}{2} - \frac{\varphi_2}{2}\right)}\,.	
\end{aligned}
\end{equation}

Having formulated the gauge theory and its gravity duals, we can proceed with the study of their deformations. General deformations of backgrounds generated by M2-branes, and their interpretation from the point of view of the dual Bagger-Lambert theory, have been derived and carefully studied in~\cite{Berman:2007tf,Berman:2008be}. Here we will mostly concentrate on the type IIA solution~\eqref{ads4cp3} and interpret the results for the dual ABJM theory. As in the case of \Ne{4} SYM, the case we will study with the most care is the $\beta$-deformation.

\subsection{$\beta$-deformation}\label{s:M2beta}

Let us study the $\beta$-deformation of the ABJM theory. The obvious isometry directions that we can use to perform the TsT transformation are $\psi$, $\varphi_1$ and $\varphi_2$. We choose the latter two, while the remaining $\psi$ isometry will correspond to the $U(1) = SO(2)$ R-symmetry of the \Ne{2} supersymmetry (in three dimensions) preserved by the deformed theory.

%: Table: ABJM charges
\TABLE{
\begin{tabular}{|c|c|c|c|c|}
\hline
& $A_1$ & $A_2$ & $B_1$ & $B_2$ \\
\hline
$U(1)_{\varphi_1}$ & $+\frac{1}{2}$ & $-\frac{1}{2}$ & $0$ & $0$ \\
\hline
$U(1)_{\varphi_2}$ & $0$ & $0$ & $-\frac{1}{2}$ & $+\frac{1}{2}$ \\
\hline
\end{tabular}
\caption{$U(1) \times U(1)$ charges of the chiral fields of the ABJM theory.\label{t:M2}}}
The charges of the fields $A_i$ and $B_i$ appearing in the superpotential~\eqref{M2W} under the two $U(1)$ factors we have chosen can be derived from~\eqref{s7emb} and are shown in table~\ref{t:M2}. Using the corresponding deformed product~\eqref{starbeta} in the case at hand, we see that the deformation results in the following modification of the superpotential:
\begin{equation}\label{M2Wbeta}
	W \to W_{\gamma} = \frac{4 \pi}{k} \tr \left( e^{- i \pi \gamma / 2} A_1 B_1 A_2 B_2
		- e^{i \pi \gamma / 2} A_1 B_2 A_2 B_1 \right)\,.
\end{equation}
As in the case of \Ne{4} SYM that we have studied in subsection~\ref{s:SYMbeta}, the $\beta$-deformed ABJM theory has a richer structure of vacua than its undeformed counterpart, as we will show in the following by making use of D-brane probes in the gravity dual. 

Let us then apply the \tst{\varphi_1}{\varphi_2}{\gamma} transformation to the $AdS_4 \times \cp$ background~\eqref{ads4cp3} of type IIA supergravity. After some computation, one gets the following solution, that contains an $AdS_4$ factor and a deformed \cp{} factor:
\begin{equation}\label{M2beta}
\begin{aligned}
	ds^2_{\text{IIA}} &= \frac{R^3}{k} \left( \frac{1}{4} ds^2_{AdS_4} + ds^2_{\widetilde{\cp}} \right)\,,\\
	ds^2_{\widetilde{\cp}} &= d\xi^2 + \M \cos^2 \xi \sin^2 \xi
		\left( d\psi + \frac{1}{2} \cos \theta_1 d\varphi_1 - \frac{1}{2} \cos \theta_2 d\varphi_2 \right)^2\\
		&\qquad + \frac{1}{4} \cos^2 \xi (d\theta_1^2 + \M \sin^2 \theta_1 d\varphi_1^2 )
		 + \frac{1}{4} \sin^2 \xi (d\theta_2^2 + \M \sin^2 \theta_2 d\varphi_2^2 )\\
		 &\qquad + \hgamma^2 \M \cos^4 \xi \sin^4 \xi \sin^2 \theta_1 \sin^2 \theta_2 d\psi^2\,,\\
	e^{2 \Phi} &= \frac{R^3}{k^3} \M\,,\\
	B &=  - \frac{\hgamma \M R^3}{k} \cos^2 \xi \sin^2 \xi\\
		&\qquad \times \Big( \frac{1}{2} \cos^2 \xi \sin^2 \theta_1 \cos \theta_2 d \psi \wedge d\varphi_1
		 + \frac{1}{2} \sin^2 \xi \sin^2 \theta_2 \cos \theta_1 d \psi \wedge d\varphi_2\\
		 &\qquad + \frac{1}{4} \left(\sin^2 \theta_1 \sin^2 \theta_2
		+ \cos^2 \xi \sin^2 \theta_1 \cos^2 \theta_2
		+ \sin^2 \xi \sin^2 \theta_2 \cos^2 \theta_1\right) d \varphi_1 \wedge d \varphi_2 \Big)\,,\\
	F_2 &= k  \Big( - \cos \xi \sin \xi d\xi
		\wedge (2 d\psi + \cos \theta_1 d \varphi_1 - \cos \theta_2 d \varphi_2)\\
		&\qquad -\frac{1}{2} \cos^2\xi \sin\theta_1 d\theta_1 \wedge d\varphi_1
		-\frac{1}{2} \sin^2\xi \sin\theta_2 d\theta_2 \wedge d\varphi_2\Big)\,,\\
	F_4 &= - \frac{3 R^3}{8} \left( \omega_{AdS_4}
		+ 4 \hgamma \cos^3 \xi \sin^3 \xi \sin \theta_1 \sin \theta_2
			d\xi \wedge d\psi \wedge d \theta_1 \wedge d \theta_2 \right)\\
		&\qquad - \frac{R^3}{8} d (\hgamma \M \cos^2 \xi \sin^2 \xi
			( \cos^2 \xi \sin^2 \theta_1 - \sin^2 \xi \sin^2 \theta_2 ) )
			\wedge d\psi \wedge d\varphi_1 \wedge d\varphi_2 \,,
\end{aligned}
\end{equation}
where $\hgamma = \frac{R^3}{4 k} \gamma$ and:
\begin{equation}\label{M2M}
	\M^{-1} = 1 + \hgamma^2 \cos^2 \xi \sin^2 \xi
		\left( \sin^2 \theta_1 \sin^2 \theta_2
		+ \cos^2 \xi \sin^2 \theta_1 \cos^2 \theta_2
		+ \sin^2 \xi \sin^2 \theta_2 \cos^2 \theta_1 \right)\,.
\end{equation}
As in the case of $\beta$-deformed \Ne{4} SYM, we see that the fact that the deformation is marginal translates in the gravity solution into the fact that the $AdS_4$ factor is untouched. The solution~\eqref{M2beta} can be seen to preserve four supercharges, matching the dual \Ne{2} supersymmetric three-dimensional gauge theory.

If we are in the range of parameter where the appropriate gravity dual of the $\beta$-deformed Chern-Simons theory lives in eleven dimensional supergravity rather than in type IIA, we can compute the eleven dimensional uplift of~\eqref{M2beta} that reads:
\begin{equation}\label{M2beta11}
\begin{aligned}
	ds^2_{11} &= \M^{-1/3} \left( \frac{R^2}{4} ds^2_{AdS_4} + R^2 ds^2_{\widetilde{\sk}} \right)\,,\\
	ds^2_{\widetilde{\sk}} &= ds^2_{\widetilde{\cp}} + \frac{ \M}{k^2} ( dy + C_1 )^2\,,\\
	G_4 &= - \frac{3 R^3}{8} \left( \omega_{AdS_4}
		+ 4 \hgamma \cos^3 \xi \sin^3 \xi \sin \theta_1 \sin \theta_2
			d\xi \wedge d\psi \wedge d \theta_1 \wedge d \theta_2 \right)\\
		&\qquad + d ( B \wedge ( dy + C_1 ) )\,,
\end{aligned}
\end{equation}
where $ds^2_{\widetilde{\cp}}$, $C_1$, $B$ and $\M$ are quantities of the type IIA solution~\eqref{M2beta}. Notice that $\beta$-type deformations of non-orbifolded $AdS_4 \times S^7$ were constructed and analyzed in~\cite{Lunin:2005jy,Berman:2007tf,Berman:2008be}.

Having derived the deformed solutions, we now try to elucidate some of their structure by using D-brane probes, as we did in subsection~\ref{s:SYMbeta} for the case of $\beta$-deformed \Ne{4} SYM. The moduli space of the undeformed theory can be analyzed by means of D2-brane probes in type IIA, or M2-brane probes in M-theory.

From our general observations in sections~\ref{s:open} and~\ref{s:branes} and from the explicit example in section~\ref{s:SYM}, by now we know very well that a D2-brane probe embedded along $x^{a}$, $a=0,1,2$, which is the only allowed static configuration when $\gamma$ is not rational, will feel a static potential that will force the brane to sit at one of the points where the torus $( \varphi_1 , \varphi_2 )$ shrinks and $\M = 1$: 
\begin{equation}\label{6cases}
\begin{gathered}
	\text{(i)} \quad \xi = 0\,,\qquad
	\text{(ii)} \quad \xi = \pi/2\,,\qquad
	\text{(iii)} \quad \theta_1 = \theta_2 = 0\,,\qquad
	\text{(iv)} \quad \theta_1 = \theta_2 = \pi\,, \\
	\text{(v)} \quad \theta_1 = 0\,,\quad \theta_2 = \pi\,,\qquad
	\text{(vi)} \quad \theta_1 = \pi\,, \quad \theta_2 = 0\,.
\end{gathered}
\end{equation}
Let us consider case (i) in some detail. As usual, we embed the brane in the static gauge, then fix $\xi = 0$. We also allow for a non-zero gauge field strength $F_{ab}$ on the world-volume, the reason being that, if our aim is to get the full moduli space of the theory, we will need to dualize the three-dimensional gauge field into an additional scalar. The D2-brane action:
\begin{equation}
	S_{\text{D}2} = - \tau_2 \int d^{3}\sigma\ e^{-\Phi}
		\sqrt{- \det \left(\hat{G}_{ab} + \hat{B}_{ab} + F_{ab}\right) }
		 + \tau_2 \int ( \hat{C}_3 + \hat{C}_1 \wedge ( \hat{B} + F) )\,,
\end{equation}
with the choices we have made reduces to:
\begin{multline}\label{R?}
	S_{\text{D}2}^{\text{TsT}} = - \frac{\tau_2 R^3}{8} \int d^{3}\sigma\  r^3 \Bigg[ \sqrt{
		1 + \frac{1}{r^4} \left( ( \partial_a r)^2 + r^2 \left( \tfrac{8 k^2}{R^6} (F_{ab})^2
		+ ( \partial_a \theta_1)^2
		+ \sin^2 \theta_1 ( \partial_a \varphi_1)^2 \right) \right)}\\
		- 1 - \frac{2 k}{R^3 r^3}
			\epsilon^{abc} \cos \theta_1 \partial_a \varphi_1 F_{bc} \Bigg]\,.
\end{multline}
We now expand the square root up to quadratic order in derivatives, and introduce a Lagrange multiplier:
\begin{equation}\label{lagr}
	\tau_2 \int y\ dF\,,
\end{equation}
so that the equation of motion for $y$ enforces the Bianchi identity for $F$ on-shell. We can now regard the action as a function of the field-strength and, after integrating the term~\eqref{lagr} by parts, integrate $F$ out by using its equation of motion. The result reads:
\begin{equation}
	S_{\text{D}2}^{\text{TsT}} = - \frac{\tau_2 R^3}{16} \int d^{3}\sigma \left[
		\frac{( \partial_a r)^2}{r}
		+ r \left( ( \partial_a \theta_1)^2
		+ ( \partial_a \varphi_1)^2
		+ \left( \tfrac{2 \partial_a y}{k} \right)^2
		+ 2 \cos \theta_1 \partial_a \varphi_1 \left( \tfrac{2 \partial_a y}{k} \right)
		\right) \right]\,,
\end{equation}
that with the change of coordinates $\tfrac{2 y}{k} = \alpha$, $r = \rho^2$ becomes:
\begin{equation}\label{S3}
	S_{\text{D}2}^{\text{TsT}} = - \frac{\tau_2 R^3}{4} \int d^{3}\sigma \left[
		( \partial_a \rho)^2
		+ \frac{\rho^2}{4} \left( ( \partial_a \theta_1)^2
		+ ( \partial_a \varphi_1)^2
		+ ( \partial_a \alpha )^2
		+ 2 \cos \theta_1 \partial_a \varphi_1 \partial_a \alpha 
		\right) \right]\,.
\end{equation}
The metric in the $( \theta_1, \varphi_1, \alpha)$ space is the one of a round three-sphere expressed in terms of Euler angles. However, the periodicity of the angle $\alpha$ is $4 \pi / k$ instead of the $4 \pi$ required for a sphere, so the full moduli space read by the world-volume action of our D2-brane probe is the orbifold $\mathbb{R}^4 / \mathbb{Z}_k$.

The computation of the analogous D2-brane probe in any of the other cases in~\eqref{6cases}  precisely reduces to~\eqref{S3}. For instance, if we choose case (iii) with $\theta_1 = \theta_2 = 0$, the final result becomes identical to~\eqref{S3} if we make the replacement $\xi \to \tfrac{\theta_1}{2}$, $ ( \psi + \tfrac{\varphi_1}{2} - \tfrac{\varphi_2}{2} ) \to \varphi_1$. We then conclude that the moduli space of the abelian $\beta$-deformed ABJM theory, for generic values of the deformation parameter $\gamma$, is made up of six copies of $\mathbb{R}^4 / \mathbb{Z}_k$.

This is in agreement with gauge theory expectations, since the F-term equations arising from the $\beta$-deformed superpotential~\eqref{M2Wbeta},
\begin{equation}
\begin{aligned}
	B_1 A_2 B_2 - e^{i \pi \gamma} B_2 A_2 B_1 &= 0\,, & 
	B_2 A_1 B_1 - e^{i \pi \gamma} B_1 A_1 B_2 &= 0\,,\\
	A_2 B_2 A_1 - e^{i \pi \gamma} A_1 B_2 A_2 &= 0\,, &
	A_1 B_1 A_2 - e^{i \pi \gamma} A_2 B_1 A_1 &= 0\,,	
\end{aligned}
\end{equation}
for $\gamma \neq 0$ are solved by setting two out of the four fields $A_i$ and $B_i$ to zero. As we can see from~\eqref{s7emb}, each one of the six possibilities corresponds to one the six cases in~\eqref{6cases} and spans an $\mathbb{R}^4 / \mathbb{Z}_k$ space. As usual, the nonabelian moduli space will be obtained from the abelian one by means of a symmetrized product.

We expect new branches to arise when $\gamma$ is rational in the $\beta$-deformed ABJM theory under consideration too. As we are by now used to see, on the string side we should then be able to find a D4-brane wrapped on the two-torus $(\varphi_1, \varphi_2)$ with a flux $F_{\varphi_1\varphi_2} = 1/\gamma$ turned on. Putting together all we have learnt until now, we then study a D4-brane probe embedded as:
\begin{equation}
	x^a = \sigma^a\ (a=0,1,2)\,,\quad
	\varphi_1 = \sigma^3\,,\quad
	\varphi_2 = \sigma^4\,,
\end{equation}
with all the other coordinates depending on $x^a$, and with a world-volume field strength:
\begin{equation}
	F_{ab}\,,\qquad
	F_{34} = \frac{1}{\gamma}\,,\qquad
	F_{a3} = \frac{1}{\gamma} \partial_a \varphi_2 (x^a) \,,\qquad
	F_{a4} = - \frac{1}{\gamma} \partial_a \varphi_1 (x^a) \,.
\end{equation}
The parameterization of $F$ has been chosen in such a way that we have already traded the Wilson lines on the torus with periodic scalars $\varphi_i$, with period $2 \pi / n$, as in~\eqref{ident} and~\eqref{varphiA}. The D4-brane action:
\begin{multline}\label{D4}
	S_{\text{D}4} = - \tau_4 \int d^{5}\sigma\ e^{-\Phi}
		\sqrt{- \det \left(\hat{G}_{ab} + \hat{B}_{ab} + F_{ab}\right) }\\
		 + \tau_4 \int \left( \hat{C}_5 + \hat{C}_3 \wedge ( \hat{B} + F )
		  + \tfrac{1}{2} \hat{C}_1 \wedge ( \hat{B} + F ) \wedge ( \hat{B} + F ) \right) \,,
\end{multline}
when we expand the square root, perform the change of coordinates $r = \rho^2$ and integrate over $\sigma^3$ and $\sigma^4$, becomes:
\begin{equation}\label{D4TsT}
\begin{split}
	S_{\text{D}4}^{\text{TsT}} &= - \frac{1}{\gamma} \frac{\tau_2 R^3}{4} \int d^{3}\sigma \bigg[ 
		(\partial_a \rho)^2 + \rho^2 \bigg( \tfrac{2 k^2}{R^6} (F_{ab})^2 + (\partial_a \xi)^2\\
		&\quad + \cos^2 \xi \sin^2 \xi \left( \partial_a \psi 
		+ \tfrac{1}{2} \cos \theta_1 \partial_a \varphi_1
		- \tfrac{1}{2} \cos \theta_2 \partial_a \varphi_2 \right)^2\\
		&\quad + \tfrac{1}{4} \cos^2 \xi \left( (\partial_a \theta_1)^2
			+ \sin^2 \theta_1(\partial_a \varphi_1)^2 \right)
		+ \tfrac{1}{4} \sin^2 \xi \left( (\partial_a \theta_2)^2
			+ \sin^2 \theta_2 (\partial_a \varphi_2)^2 \right)\\
		&\quad + \frac{4}{R^3 r^3} \epsilon^{abc} ( C_1)_a F_{bc}
	\bigg) \bigg]\,,
\end{split}
\end{equation}
which is the same action we would have obtained for a D2-brane in the undeformed ABJM factor divided by a factor of $\gamma$, $S_{\text{D}4}^{\text{TsT}} = S_{\text{D}2}^{(0)} / \gamma$. In order to obtain the full moduli space, we need to dualize the three-dimensional gauge field onto a periodic scalar, as we did above in the case of the D2-brane. The result reads:
\begin{equation}\label{Mmoduli}
\begin{split}
	S_{\text{D}4}^{\text{TsT}} &= - \frac{1}{\gamma} \frac{\tau_2 R^3}{4} \int d^{3}\sigma \bigg[ 
		(\partial_a \rho)^2 + \rho^2 \bigg( (\partial_a \xi)^2\\
		&\quad + \cos^2 \xi \sin^2 \xi \left( \partial_a \psi
		+ \tfrac{1}{2} \cos \theta_1 \partial_a \varphi_1
		- \tfrac{1}{2} \cos \theta_2 \partial_a \varphi_2 \right)^2\\
		&\quad + \tfrac{1}{4} \cos^2 \xi \left( (\partial_a \theta_1)^2
			+ \sin^2 \theta_1(\partial_a \varphi_1)^2 \right)
		+ \tfrac{1}{4} \sin^2 \xi \left( (\partial_a \theta_2)^2
			+ \sin^2 \theta_2 (\partial_a \varphi_2)^2 \right)\\
		&\quad +\frac{1}{k^2} \left( \partial_a y + (C_1)_\psi \partial_a \psi
			 + (C_1)_{\varphi_1} \partial_a \varphi_1
			 + (C_1)_{\varphi_2} \partial_a \varphi_2 \right)^2
	\bigg) \bigg]\,.
\end{split}
\end{equation}
The action~\eqref{Mmoduli} gives us the metric on the additional branches of the moduli space of the $\beta$-deformed ABJM theory that arise when $\gamma$ is rational. At first sight, the metric looks like describing the space $\cZ{4}{k}$ (realized as a cone over $\sk$, see~\eqref{ads4s7k}), namely the moduli space of the abelian undeformed theory. However, as in the other cases we have considered so far, we have to take into account the new periodicities of $\varphi_1$ and $\varphi_2$. The moduli space in these additional branches will then be, once again, a $\mathbb{Z}_n \times \mathbb{Z}_n$ orbifold of the moduli space of the undeformed theory.

It would be instructive to repeat the analysis in M-theory. In fact, an M2-brane probe computation goes through precisely as the D2-brane probe computation we did above, yielding the same result (of course without the need of any dualization). The branches of the moduli space that emerge for rational $\gamma$ would instead be described by M5-branes that wrap the three-torus $(\varphi_1, \varphi_2, y)$ with a self-dual two-form turned on on their world-volume. However, we will not enter into the details of this computation here, due to the known difficulties with the M5-brane action, see for instance~\cite{Berman:2007bv} and references therein.

\subsection{3-parameter deformation}

We can generalize the gravity dual of the $\beta$-deformed three dimensional Chern-Simons-matter theory that we have found in the previous subsection by finding a three-parameter family of non-supersymmetric deformations, as we did in subsection~\eqref{s:SYMbeta3} for \ads. We then perform three successive TsT transformation, \tst{\varphi_1}{\varphi_2}{\gamma_3}, \tst{\varphi_2}{\psi}{\gamma_1} and \tst{\psi}{\varphi_1}{\gamma_2}, and the final result reads:
\begin{equation}\label{M2beta3}
\begin{aligned}
	ds^2_{\text{IIA}} &= \frac{R^3}{k} \left( \frac{1}{4} ds^2_{AdS_4} + ds^2_{\widetilde{\cp}} \right)\,,\\
	ds^2_{\widetilde{\cp}} &= d\xi^2 + \M \cos^2 \xi \sin^2 \xi
		\left( d\psi + \frac{1}{2} \cos \theta_1 d\varphi_1 - \frac{1}{2} \cos \theta_2 d\varphi_2 \right)^2\\
		&\qquad + \frac{1}{4} \cos^2 \xi (d\theta_1^2 + \M \sin^2 \theta_1 d\varphi_1^2 )
		 + \frac{1}{4} \sin^2 \xi (d\theta_2^2 + \M \sin^2 \theta_2 d\varphi_2^2 )\\
		&\qquad + \M \cos^4 \xi \sin^4 \xi \sin^2 \theta_1 \sin^2 \theta_2
			(\hgamma_1 d\varphi_1 + \hgamma_2 d\varphi_2 + \hgamma_3 d\psi )^2\,,\\
	e^{2 \Phi} &= \frac{R^3}{k^3} \M\,,\\
	B &=  - \frac{\M R^3}{k} \cos^2 \xi \sin^2 \xi
		\Big[ \frac{1}{2} (\hgamma_3 \cos \theta_2 + 2 \hgamma_2)
			\cos^2 \xi \sin^2 \theta_1 d \psi \wedge d\varphi_1\\
		&\qquad  + \frac{1}{2} (\hgamma_3 \cos \theta_1 - 2 \hgamma_1)
		 	\sin^2 \xi \sin^2 \theta_2 d \psi \wedge d\varphi_2\\
		 &\qquad + \frac{1}{4} \Big(\hgamma_3 \sin^2 \theta_1 \sin^2 \theta_2
		+ (\hgamma_3 \cos \theta_2 + 2 \hgamma_2) \cos^2 \xi \sin^2 \theta_1 \cos \theta_2\\
		&\qquad\qquad
			+ (\hgamma_3 \cos \theta_1 - 2 \hgamma_1)\sin^2 \xi \sin^2 \theta_2 \cos \theta_1\Big)
			d \varphi_1 \wedge d \varphi_2 \Big]\,,\\
	F_2 &= k  \Big( - \cos \xi \sin \xi d\xi
		\wedge (2 d\psi + \cos \theta_1 d \varphi_1 - \cos \theta_2 d \varphi_2)\\
		&\qquad -\frac{1}{2} \cos^2\xi \sin\theta_1 d\theta_1 \wedge d\varphi_1
		-\frac{1}{2} \sin^2\xi \sin\theta_2 d\theta_2 \wedge d\varphi_2\Big)\,,\\
	F_4 &= - \frac{3 R^3}{8} \big( \omega_{AdS_4}\\
		&\qquad + 4 \cos^3 \xi \sin^3 \xi \sin \theta_1 \sin \theta_2
			d\xi \wedge d \theta_1 \wedge d \theta_2
			 \wedge (\hgamma_1 d\varphi_1 + \hgamma_2 d\varphi_2 + \hgamma_3 d\psi ) \big)\\
		&\qquad - \frac{R^3}{8} d ( \M \cos^2 \xi \sin^2 \xi
			( (\hgamma_3 + 2 \hgamma_2 \cos \theta_2) \cos^2 \xi \sin^2 \theta_1\\
		&\qquad\qquad - (\hgamma_3 - 2 \hgamma_1 \cos \theta_1) \sin^2 \xi \sin^2 \theta_2 ) )
			\wedge d\psi \wedge d\varphi_1 \wedge d\varphi_2 \,,
\end{aligned}
\end{equation}
where:
\begin{multline}\label{M2M3}
	\M^{-1} = 1 + \cos^2 \xi \sin^2 \xi
		\Big( \hgamma_3^2 \sin^2 \theta_1 \sin^2 \theta_2\\
		+ (\hgamma_3 \cos \theta_2 + 2 \hgamma_2)^2 \cos^2 \xi \sin^2 \theta_1
		+ (\hgamma_3 \cos \theta_1 - 2 \hgamma_1)^2 \sin^2 \xi \sin^2 \theta_2 \Big)\,.
\end{multline}
Of course, this non-supersymmetric deformation reduces to the supersymmetric one studied in the previous subsection by putting $\hgamma_1 = \hgamma_2 = 0$ and $\hgamma_3 = \hgamma$. We can also write the eleven-dimensional uplift in complete analogy with~\eqref{M2beta11}.

\subsection{Non-commutative deformation}\label{s:M2nc}

We will not spend many words on the non-commutative deformation of the ABJM three-dimensional theory, since, on the string side, it only involves the $AdS$ directions and is thus very similar to the deformation~\eqref{N=4nc} of \ads{}. We can proceed in the same way as in subsection~\ref{s:SYMnc} to find the solution: 
\begin{equation}\label{M2nc}
\begin{aligned}
	ds^2_{\text{IIA}} &= \frac{R^3}{4 k} \left( r^2 (- (dx^0)^2 + \M ( (dx^1)^2 + (dx^2)^2)
		+ \frac{dr^2}{r^2} + 4 ds^2_{\cp} \right) \,, \\
	e^{2 \Phi} &= \frac{R^3}{k^3} \M\,, \\
	B &= - \frac{\hgamma \M R^3}{4 k} r^4 dx^1 \wedge dx^2\,,\\
	C_1 &= \frac{k}{2} \left( ( \cos^2 \xi - \sin^2 \xi ) d\psi
		+ \cos^2 \xi \cos \theta_1 d \varphi_1
		+ \sin^2 \xi \cos \theta_2 d \varphi_2 \right)
		- \frac{\hgamma R r^3}{8} dx^0\,,\\
	C_3 &= \frac{R^3 r^3 \M}{8} dx^1 \wedge dx^2
		\wedge \Big( dx^0 + \\
		&\qquad + \hgamma r  
		\left( ( \cos^2 \xi - \sin^2 \xi ) d\psi
		+ \cos^2 \xi \cos \theta_1 d \varphi_1
		+ \sin^2 \xi \cos \theta_2 d \varphi_2 \right) \Big)\,,
\end{aligned}
\end{equation}
where $ds^2_{\cp}$ is given in~\eqref{ads4cp3}, and we have defined $\hgamma = \frac{R^3}{4 k} \gamma$ and:
\begin{equation}
	\M^{-1} = 1 + \hgamma^2 r^4\,.
\end{equation}
The solution~\eqref{M2nc} will be dual to the ABJM Chern-Simons-matter theory put on a non-commutative torus $(x^1,x^2)$ with:
\begin{equation}
	[ x^1 , x^2 ] = i \theta^{12} = - 2 \pi i \gamma\,.
\end{equation}

\subsection{Dipole deformation}\label{s:M2dipole}

As we did in the case of \ads, in order to get the gravity dual of a dipole deformation of our \Ne{6} theory we must choose one $AdS_4$ and one \cp{} direction to perform the TsT. The latter direction can be a combination of the three angles $\psi$, $\varphi_1$ and $\varphi_2$ but let us just choose $\psi$ for simplicity, and perform a \tst{x^2}{\psi}{\gamma} transformation. Again, more general backgrounds can be constructed, see for instance the eleven-dimensional solutions of~\cite{Berman:2007tf}.

The resulting type IIA solution, dual to the dipole theory under consideration is:
\begin{equation}\label{M2dipole}
\begin{aligned}
	ds^2_{\text{IIA}} &= \frac{R^3}{4 k} \left( r^2 (- (dx^0)^2 + (dx^1)^2 + \M (dx^2)^2 )
		+ \frac{dr^2}{r^2} \right) \\
		&\qquad +\frac{R^3}{k} \bigg( d\xi^2 + \M \cos^2 \xi \sin^2 \xi
		\left( d\psi + \frac{1}{2} \cos \theta_1 d\varphi_1 - \frac{1}{2} \cos \theta_2 d\varphi_2 \right)^2\\
		&\qquad\qquad + \frac{1}{4} \cos^2 \xi (d\theta_1^2 + \sin^2 \theta_1 d\varphi_1^2 )
		 + \frac{1}{4} \sin^2 \xi (d\theta_2^2 + \sin^2 \theta_2 d\varphi_2^2 ) \bigg)\\
	e^{2 \Phi} &= \frac{R^3}{k^3} \M\,, \\
	B &= - \frac{\hgamma \M R^3}{k} r^2 \cos^2 \xi \sin^2 \xi dx^2
		\wedge \left( d\psi + \frac{1}{2} \cos\theta_1 d\varphi_1
		 - \frac{1}{2} \cos\theta_2 d\varphi_2 \right) \,,\\
	C_1 &= \frac{k}{2} \left( ( \cos^2 \xi - \sin^2 \xi ) d\psi
		+ \cos^2 \xi \cos \theta_1 d \varphi_1
		+ \sin^2 \xi \cos \theta_2 d \varphi_2 \right)\,,\\
	C_3 &= \frac{R^3 r^3}{8} dx^2 \wedge \Big( dx^0 \wedge dx^1 \\
		&\qquad + \frac{2 \hgamma \M}{r} \cos^2 \xi \sin^2 \xi
		\left( d \psi \wedge (\cos\theta_1 d\varphi_1 + \cos\theta_2 d\varphi_2)
		+ \cos\theta_1 \cos\theta_2 d\varphi_1 \wedge d\varphi_2\right) \Big)\,,
\end{aligned}
\end{equation}
where $\hgamma = \frac{R^3}{4 k} \gamma$ and:
\begin{equation}
	\M^{-1} = 1 + 4 \hgamma^2 r^2 \cos^2 \xi  \sin^2 \xi \,.
\end{equation}

This concludes our study of deformations of the \Ne{6} ABJM Chern-Simons-matter theory.

\acknowledgments

It is a pleasure to thank Riccardo Argurio, Sujay Ashok, Francesco Bigazzi, Frank Ferrari, Chethan Krishnan, Stanislav Kuperstein, Carlo Maccaferri, Asad Naqvi, Nemani Suryanarayana and in particular Carlos N\'u\~nez for advice and many illuminating discussions. This work is supported by the Belgian Fonds de la Recherche Fondamentale Collective (grant 2.4655.07). It is also supported in part by the Belgian Institut Interuniversitaire des Sciences Nucl\'eaires (grant 4.4505.86), the Interuniversity Attraction Poles Programme (Belgian Science Policy) and the European Commission FP6 programme MRTN-CT-2004-005104 (in association with V. U. Brussels).

\appendix

\section{Conventions and useful formulae}\label{s:conventions}

\subsection{Supergravity fields}

This appendix details the notation and conventions used throughout the paper. Notice that in this appendix we keep the string length $\ls$ explicit for completeness, while in the main text we work in units where $\ls=1$. Let us start with the bosonic part of the type IIB and type IIA supergravity actions in the string frame. The common NS-NS part reads:
\begin{equation}
	S_{\text{II}}^{\text{(NS)}} = \frac{1}{2\kappa^2} \int d^{10}x \sqrt{-\det G}\ e^{-2\Phi}R
		- \frac{1}{4 \kappa^2} \int \left[ - 8 e^{-2\Phi} d\Phi \wedge \hd{} d\Phi
		+ e^{-2\Phi}  H \wedge \hd{}H \right]\,,
\end{equation}
where $\kappa = 8 \pi^{7/2} \gs \ls^4$ and $H=dB$, while the R-R parts read respectively:
\begin{align}
	S_{\text{IIB}}^{\text{(R)}} &= - \frac{1}{4 \kappa^2} \int
		\left[ F_1 \wedge \hd{}F_1 + \mathcal{F}_3 \wedge \hd{}\mathcal{F}_3
		+ \frac{1}{2} \mathcal{F}_5 \wedge \hd{}\mathcal{F}_5
		- C_4\wedge H \wedge F_3 \right]\,,\\
	S_{\text{IIA}}^{\text{(R)}} &= \frac{1}{4 \kappa^2} \int
		\left[ F_2 \wedge \hd{}F_2 + \mathcal{F}_4 \wedge \hd{}\mathcal{F}_4
		- B \wedge F_4 \wedge F_4 \right]\,,
\end{align}
where $F_p = dC_{p-1}$ and the modified field strengths $\mathcal{F}_p$ are defined as:
\begin{equation}\label{modifiedF}
	\mathcal{F}_p = F_p + H \wedge C_{p-3}\,.
\end{equation}
Higher rank forms with $p>5$ are defined via Hodge duality, $\mathcal{F}_p = \hd{} \mathcal{F}_{10-p}$. The self-duality of the five-form, $\mathcal{F}_5 = \hd{} \mathcal{F}_5$, is not taken into account by the type IIB action and has to be imposed on-shell.

The action of eleven-dimensional supergravity is given by:
\begin{equation}
	S_{11} = \frac{1}{2\kappa_{11}^2} \int d^{11}x \sqrt{-\det G}\ R
		+ \frac{1}{4 \kappa^2} \int
		\left[ G_4 \wedge \hd{}G_4
		- \frac{1}{3} A_3 \wedge G_4 \wedge G_4 \right]\,,
\end{equation}
where $G_4 = dA_3$ and $\kappa_{11} = 2^{7/2} \pi^4 \lp^{9/2}$, $\lp$ being the eleven-dimensional Planck length. If we compactify eleven-dimensional supergravity on a circle $x_{10}$ of radius $R_{11}$, we get type IIA supergravity with the identifications:
\begin{equation}
	R_{11} = \gs \ls\,,\qquad \lp = \gs^{1/3} \ls\,.
\end{equation}
The reduction ansatz for the fields is given by:
\begin{equation}
\begin{split}
	ds^2_{11} &= e^{-2 \Phi/3} ds^2_{\text{IIA}} + e^{4 \Phi/3} (d x_{10} + C_1)^2\,,\\
	G_4 &= F_4 + H_3 \wedge d x_{10} \,.
\end{split}
\end{equation}

\subsection{T-duality}

We now summarize the T-duality rules, starting with background fields. Denote the direction along which the T-duality acts by $y$ and the remaining directions by $\alpha,\beta,\ldots$, and assume that no field depends on $y$. Metric and dilaton transform as:
\begin{equation}
	G'_{yy} = \frac{1}{G_{yy}}\,,\quad
	G'_{\alpha y} = \frac{B_{\alpha y}}{G_{yy}}\,,\quad
	G'_{\alpha \beta} = G_{\alpha \beta}
		- \frac{G_{\alpha y} G_{\beta y} - B_{\alpha y} B_{\beta y}}{G_{yy}}\,,\quad
	e^{2 \Phi'} = \frac{e^{2 \Phi}}{G_{yy}}\,,
\end{equation}
where primed fields are the ones after the transformation (the duality of course exchanges type IIA and IIB). We want to write the transformations of the NS-NS and R-R antisymmetric tensor fields in terms of differential forms. Given a $p$-form $\omega_p$, we decompose it as:
\begin{equation}
	\omega_p = \bar{\omega}_p + \omega_{p [y]} \wedge dy\,,
\end{equation}
where $\bar{\omega}_p = \frac{1}{p!} \omega_{\alpha_1 \dotsm \alpha_p} dx^{\alpha_1} \wedge \dotsm \wedge dx^{\alpha_p}$ does not contain any $dy$ components, while $\omega_{p [y]}$ is a $(p-1)$-form whose components are given by:%
\footnote{This is of course just a short-hand notation for the interior product $\iota$, an anticommuting operator of form degree $-1$.}
\begin{equation}\label{omegay}
	(\omega_{p [y]})_{\alpha_1 \dotsm \alpha_{p-1}} = (\omega_p)_{\alpha_1 \dotsm \alpha_{p-1} y}\,.
\end{equation}
Finally, we define two one-forms, $j$ and $b$, as:
\begin{equation}
	j = \frac{G_{\alpha y}}{G_{yy}} dx^{\alpha}\,,\qquad
	b = B_{[y]} + dy\,.
\end{equation}
The T-duality rules for the NS-NS and R-R potentials are then given by:
\begin{equation}\label{Tdualpot}
\begin{split}
	B' &= \bar{B} + j \wedge b\,,\\
	C'_p &= C_{p+1[y]} + \bar{C}_{p-1} \wedge b + C_{p-1 [y]} \wedge b \wedge j\,.
\end{split}
\end{equation}
It is useful to express the T-duality rules in terms of $H$ and of the modified field strengths $\mathcal{F}_p$ defined in~\eqref{modifiedF}. They read:
\begin{equation}\label{TdualF}
\begin{split}
	H' &= \bar{H} + dj \wedge b - j \wedge db\,,\\
	\mathcal{F}'_p &= \mathcal{F}_{p+1[y]} + \bar{\mathcal{F}}_{p-1} \wedge b
		+ \mathcal{F}_{p-1 [y]} \wedge b \wedge j\,,
\end{split}
\end{equation}
and we also note that $db = H_{[y]}$.

At the level of the world-sheet string sigma-model, T-duality acts on the coordinate fields as follows. Denote with $X^1$ the direction along which the T-duality is performed and by $\tilde{X}^1$ the T-dual coordinate. We have:
\begin{equation}\label{wsTduality}
		\partial_{\alpha} \tilde{X}^1
			= \eta_{\alpha \beta} \epsilon^{\beta \kappa} G_{\mu 1} \partial_{\kappa} X^\mu
			- B_{\mu 1} \partial_{\alpha} X^\mu\,,
\end{equation}
where the indices $\alpha = (\tau, \sigma)$ denote the world-volume coordinates, $\eta_{\alpha \beta}$ is the world-sheet metric and the antisymmetric symbol is defined as $\epsilon^{\tau \sigma} = +1$.

\subsection{D-brane action}

The world-volume action of a D$p$-brane in the string frame which is consistent with the above supergravity actions and T-duality rules reads:
\begin{equation}\label{Dpwv}
	S_{\text{D}p} = - \tau_p \int d^{p+1}\sigma\ e^{-\Phi}
		\sqrt{- \det \left(\hat{G}_{ab} + \hat{B}_{ab} + F_{ab}\right) }
		 + \tau_p \int \sum_q \hat{C}_q \wedge e^{\hat{B} + F}\,,
\end{equation}
where $\tau_p = \frac{1}{(2\pi)^p \gs \ls^{p+1}}$, the integrals are performed on the $(p+1)$-dimensional world-volume spanned by the coordinates $\sigma^i$, hats denote pull-backs of the bulk fields onto the world-volume, and finally $F$ is the gauge field strength on the brane (we have reabsorbed a factor of $2 \pi \ls^2$ into its definition).

% Bibliography
\bibliographystyle{../bibtex/myutcaps}
\bibliography{../bibtex/mybib}

\end{document}